# Soft X-ray properties of narrow-line Seyfert 1 galaxies

Th. Boller[1], W. N. Brandt[2], H. Fink[3]

[1] Astrophysikalisches Institut Potsdam, An der Sternwarte 16, 14482 Potsdam, Germany

[2] Institute of Astronomy, Madingley Road, Cambridge CB3 0HA, U.K.

[3] Max-Planck-Institut für Extraterrestrische Physik, 85748 Garching, Germany

**Abstract.** We report on AGN with extremely soft X-ray spectra observed with ROSAT. From their optical emission lines these objects are classified as narrow-line Seyfert 1 galaxies (NLS1), almost all with extremely large Fe II/H$\beta$ ratios and relatively narrow optical lines of hydrogen. Our results are based on a systematic ROSAT study of 46 NLS1. 32 of these are located in the fields of view of ROSAT pointings and 31 are detected above a $5\sigma$ limit.

We find that NLS1 have generally steeper soft X-ray continuum slopes than normal Seyfert 1s, and there may exist an anticorrelation between 0.1–2.4 keV continuum slope and the FWHM of the H$\beta$ line. Objects with steep 0.1–2.4 keV continuum slopes and H$\beta$ FWHM > 3000 km s$^{-1}$ are clearly discriminated against by nature. When simple power-law models are fit to the data, photon indices reach values up to about 5, much higher than is usually seen in Seyfert 1s. We discuss steep ROSAT spectra in light of soft X-ray excess and hard X-ray tail models.

We do not find evidence for large neutral hydrogen column densities over the Galactic column, and thus barring very high dust-to-gas ratios simple obscuration by dust and cold gas along the line of sight cannot easily explain the narrow optical hydrogen Balmer lines of NLS1. Many NLS1 show rapid soft X-ray variability and thus significant electron scattering of their X-rays seems unlikely.

We consider models for NLS1 where they are Seyfert 1s with extremal values of pole-on orientation, black hole mass and/or accretion rate, warm absorption and BLR thickness and confront these models with the known properties of NLS1. All simple models appear to have drawbacks, but models with smaller mass black holes and thicker BLRs show some promise. We suggest specific further tests of the models. Steep ROSAT spectra suggest that the EUV and X-ray spectral distributions of NLS1 may be somewhat different than those of normal Seyfert 1s, and these different spectral energy distributions can strongly influence BLR cloud formation and confinement. NLS1 may be analogous to the high ultrasoft states of Galactic black hole candidates.

## 1. Introduction

Narrow-line Seyfert 1 galaxies (hereafter NLS1) are a peculiar group of AGNs where (1) the Balmer lines of hydrogen are only slightly broader than the forbidden lines such as [O III], [N II] and [S II]; (2) there are often present emission lines from Fe II (e.g. the optical multiplets centered at 4570 Å, 5190 Å and 5300 Å) or higher ionization iron lines such as [Fe VII] 6087 Å and [Fe X] 6375 Å (these lines are often seen in Seyfert 1 galaxies but generally not in Seyfert 2 galaxies); and (3) the ratio of [O III] 5007 Å to H$\beta$ is < 3, a level which Shuder and Osterbrock (1981) found to discriminate well Seyfert 1s from Seyfert 2s. The full width at half maximum (FWHM) of NLS1 hydrogen Balmer lines is usually in the range $\approx$ 500–1500 km s$^{-1}$. Perhaps the most thoroughly studied NLS1 is I Zw 1 (e.g. Halpern & Oke 1987). I Zw 1 shows very strong, narrow optical Fe II lines and shares other properties with broad-line AGNs, such as violent gas kinematics at the nucleus (Phillips 1978) and high X-ray luminosity (Kruper et al. 1990). NLS1 comprise of order 10 per cent of optically selected Seyfert 1s, and soft X-ray selected Seyfert 1 samples contain relatively large numbers ($\sim$16–50 per cent) of NLS1 (Stephens 1989; table 5 of Puchnarewicz et al. 1992; Sect. 6 of Puchnarewicz et al. 1995). Goodrich (1989) has taken the point of view that narrow-line Seyfert 1s are those Seyfert 1s with small line-of-sight velocity dispersions in the Broad Line Region (BLR) and has addressed this issue with spectropolarimetry. Ulvestad, Antonucci & Goodrich (1995) have reported VLA observations of seven NLS1. Only Mrk 766 was resolved in their observations, and it was found that it has its radio axis oriented approximately perpendicular to its optical polarization position angle. This is also true for the NLS1 Mrk 1126 but not for the NLS1 5C3.100. In contrast all other Seyfert 1 galaxies where both the radio axes and the optical polarization position angles have been measured have radio axes parallel to their optical polarization position angles. A perpendicular orientation is what is expected from some accretion disc models (cf. Sect. III.A of Antonucci 1992), but the presence of absorption opacity can lead to a rotation of the expected polarization angle (Matt, Fabian & Ross 1993). Ulvestad, Antonucci & Goodrich (1995) point out that NGC 1068 might well be classified as a NLS1 if it were seen along the symmetry axis of its nucleus and suggest that perhaps only the NLS1 galaxies and not all Seyfert 1s are the pole-on equivalents of Seyfert 2s.

The reason for the relatively narrow BLR line widths of NLS1 is not known. One possibility has to do with the observer's viewing angle and the geometry and dynamics of the BLR (e.g. a flattened rotating system seen pole-on). Another possibility is that BLR cloud formation in the innermost part of the BLR (where cloud velocities would be highest) is hindered by irradiation (cf. Guilbert, Fabian & McCray 1983). The part of the BLR producing the very broad component might also be obscured from direct view.

Most NLS1 show unusually strong Fe II lines relative to H$\beta$ with Fe II 4570 Å/H$\beta$ > 0.8. The median value of the distribution of this ratio in the AGN sample of Joly (1991) is about 0.5. Fe II lines are observed in Seyfert 1 galaxies, radio quiet QSOs, core dominated radio-loud QSOs and (with somewhat weaker intensity) in lobe-dominated quasars (cf. Wills, Netzer & Wills 1985 and Joly 1991 for reviews and references). These lines carry about one fourth of the total amount of energy emitted by the BLR.

The excitation mechanism of Fe II lines is one of the outstanding problems of AGN research. The two alternative models can be summarized as follows: (1) The lines are formed in the 'standard' BLR where the conditions are not too different from those leading to the formation of the other prominent emission lines, such as H$\alpha$, Ly$\alpha$ and C IV 1549Å. The ionization parameter at the Fe II zone must be low and the density and perhaps the iron abundance rather high (Wills, Netzer & Wills 1985). (2) The Fe II lines are formed in the surface layers of a accretion disk, $\sim$ 1000 gravitational radii from the central black hole. This model, which was developed by Collin-Souffrin and collaborators (e.g. Collin-Souffrin et al. 1988), depends crucially on the hard X-ray luminosity of the central source.

In this paper we report on the soft X-ray properties of NLS1. The motivation for this study and sample selection are given in Sect. 2. The X-ray observations of NLS1 and the data reduction are summarized in Sect. 3. Object notes and X-ray results for specific NLS1 are presented in Sect. 4, and results for NLS1 as a class are given in Sect. 5. Models for NLS1 and relevant physical processes are explained and confronted with the data in Sect. 6, and the summary can be found in Sect. 7.

We will sometimes speak of NLS1 and normal Seyfert 1s separately. However, it is to be understood that we use the adjective 'normal' only as a semantic discriminator and are aware both of the wide heterogeneity of Seyfert 1s as well as the fact that the properties of NLS1 and normal Seyfert 1s probably form continuous distributions (as is described below).

A value of the Hubble constant of $H_0 = 50$ km s$^{-1}$ Mpc$^{-1}$ and a cosmological deceleration parameter of $q_0 = \frac{1}{2}$ have been adopted throughout.

## 2. Motivation and sample selection

Recently Boller et al. (1993) discovered rapid X-ray variability in the NLS1 object IRAS 13224 − 3809 with a doubling time of $\approx$ 800 s (see Fig. 1). The apparent efficiency derived from this variability is constrained to be greater than about 8 per cent using the arguments of Fabian (1992). This large efficiency exceeds the maximum Schwarzschild black hole accretion efficiency of 5.7 per cent, suggesting either relativistic beaming or accretion onto a Kerr black hole (realistic accretion onto a Kerr black hole can have an accretion efficiency of up to 29 per cent; Thorne 1974).

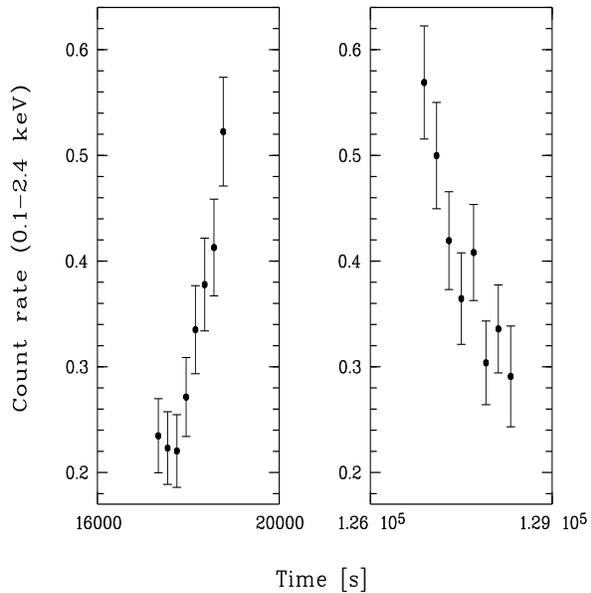

**Fig. 1.** Part of the X-ray light curve of IRAS 13224 − 3809. Amplitude variations by a factor of 2 are detected on time scales of 800 and 1000 s. The bin size is 200 s to illustrate the rapid X-ray variations. The coherent variations by a factor of 2 over integral multiples of the wobble period indicate that they are not an artifact of the wobble. The whole X-ray light curve can be found in Boller et al. (1993). The count rate is measured in counts s$^{-1}$.

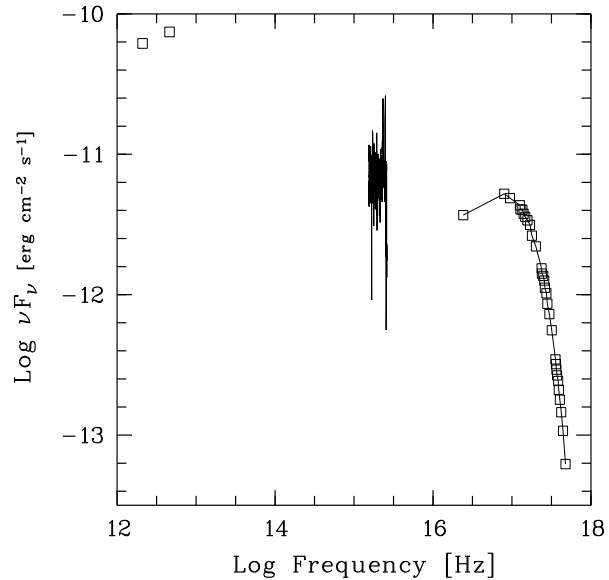

**Fig. 2.** Spectral energy distribution of IRAS 13224 − 3809. Far-infrared data are taken from the IRAS Point Source Catalogue and UV measurements are taken from Mas-Hesse et al. (1994). Note the extremely steep 0.1–2.4 keV X-ray continuum slope. Corrections for dust reddening and cold gas absorption have been made.

Fig. 2 gives the continuum energy distribution of IRAS $13224 - 3809$ from far-infrared to soft X-ray energies. It shows an extremely steep soft X-ray spectrum. A simple power-law fit to the ROSAT data gives a photon index of 4.4, much steeper than the mean value found for normal Seyfert 1s observed with ROSAT (e.g. Walter & Fink 1993). IRAS $13224 - 3809$ is also an extreme optical Fe II emitter. The optical spectrum has Fe II 4570 Å/H$\beta$ = 2.4, which makes this object one of the most intense Fe II emitters known.

We undertook a systematic investigation of other NLS1 due to the remarkable properties of IRAS $13224 - 3809$ as well as the fact that the NLS1 Ark 564 showed a steep and complex spectrum in a recent ROSAT observation (Brandt et al. 1994). We searched the literature for reports of NLS1 and found 46 in total (cf. Table 1). We did not include Mrk 1388 since Osterbrock (1985) and Goodrich (1989) state that this object is more properly classified as a Seyfert 2. 32 of these NLS1 are located in the fields of view of ROSAT pointings in the public archive. Only one (E1511+671 = MCG +11 − 19 − 003) of the 32 objects, located off-axis in the ROSAT pointings, was not found above a $5\sigma$ detection limit. The soft X-ray and optical properties of the 31 NLS1 with measured ROSAT fluxes are described in the following sections.

We point out below that several other objects (e.g. NGC 4051) that have been previously classified as normal Seyfert 1s might be profitably thought of as NLS1. We shall not include these objects in our sample since we have chosen for our sample only those objects that have been previously classified as NLS1 in the literature.

## 3. Observations and data reduction

The X-ray observations were obtained with the PSPC detector (Pfeffermann et al. 1987) on the ROSAT satellite (Trümper 1983). The PSPC has a one degree radius field of view and all observations were performed with the detector in 'wobble' mode. The accumulated time of on-source data ranges from 2.4 to 42.1 ks. The moderate Galactic neutral hydrogen column densities of $(1.0–6.2) \cdot 10^{20}$ cm$^{-2}$ (Stark et al. 1992) allow reliable spectral fits for our sources. The observation log for the NLS1 galaxies in our sample is given in Table 2.

Source extraction, background extraction and data preparation were performed with the EXSAS software package (Zimmermann et al. 1992). The spectral analysis includes spectral fitting assuming a single single power-law model with cold absorption. Variability and spectral variability studies were also performed. In some of the analysis below we have employed the ROSAT hardness ratio defined as $HR = (H − S)/(H + S)$. $H$ represents the counts in the hard (0.4–2.4 keV) energy band and $S$ the counts in the soft (0.1–0.4 keV) band. We are keenly aware of the remaining spectral calibration uncertainties with the PSPC, and refer the reader to the discussions of this issue in appendix A of Brinkmann et al. (1994), and appendix A and appendix B of Fiore et al. (1994). It is important to note, however, that the spectra of NLS1 are generally steep *even when compared to* those of other ROSAT observed Seyfert 1 galaxies.

For most sources we have binned the data in bins with widths that are an integral multiple of 400 s. We do this to avoid apparent flux variations due to the ROSAT wobble period of about 400 s. Brinkmann et al. (1994) state that flux determination in wobble mode is good to within ∼ 4 per cent when binning over integer multiples of the 400 s wobble period.

We have searched for unusual high background events due to sudden increases of charged particles. We have also looked at other sources in the fields of view to confirm that they do not show coordinated variations with those seen in the NLS1.

## 4. Object notes and specific X-ray results

In this section we present the details of the ROSAT data for the NLS1 in our sample and also point out other germane facts. Results are summarized in Table 1, and the corresponding observation log is shown in Table 2. Note that the objects are not described in order of right ascension, but are grouped according to certain distinguishing characteristics. The X-ray light curves for objects which show evidence for variability are presented in Fig. 3–6. Soft X-ray spectra for four NLS1 are shown in Fig. 7.

### 4.1. IRAS 13224−3809

Boller et al. (1993) present a detailed description and discussion of ROSAT and optical observations of IRAS $13224 - 3809$. Some information is also presented in Sect. 2.

None of the standard power-law and blackbody models gives an acceptable fit to the steep pulse-height spectrum of IRAS $13224 - 3809$. Bremsstrahlung and simple thin accretion disc models can fit the steep spectrum, but have difficulty explaining the rapid variability. Even though a simple power-law model is not a satisfactory description of the spectrum, the photon index of 4.4 serves as a rough quantitative measure of its steepness (see Fig. 2).

Using a two day ASCA observation, Otani and collaborators have found very large X-ray variability (a factor of ∼ 50 in ASCA count rate) in IRAS 13224−3809 (Otani 1995) as well as unusual spectral variability.

The X-ray position of IRAS 13224−3809 from a recent ROSAT HRI observation (Boller 1995, in preparation) is R.A.(2000)=$13^h 25^m 19.5^s$, Dec.(2000)=$-38°24'48''$. This agrees well with the improved optical position of R.A.(2000)=$13^h 25^m 19.2^s$, Dec.(2000)=$-38°24'52''$, obtained from the ROE COSMOS scans of the UK Schmidt IIIaJ southern sky survey (MacGillivray & Stobie 1985).

### 4.2. I Zw 1 (PG 0005+124)

I Zw 1, often referred to as the prototype narrow-line Seyfert 1 galaxy with Fe II emission, has been widely studied at optical wavelengths. Fig. 3 shows the ROSAT light curve and the results of a power-law fit to the ROSAT pointed data on I Zw 1 are shown in Fig. 7. We searched for spectral variability employing the hardness ratio defined in Sect. 3. The count rate interval was divided into two subsets ranging from 0.6–0.8 count s$^{-1}$ (low state) and 0.8–1.0 count s$^{-1}$ (high state). The counts in the soft and hard energy bands were determined for both subsets. The hardness ratio for the low state of $-0.26 \pm 0.04$ is not significantly higher than that for the high state of $-0.24 \pm 0.03$ (90 per cent confidence errors are used for all hardness ratios).

### 4.3. 5C3.100 (Mrk 957, IRAS 00391+4004)

5C3.100 is a I Zw 1 class object and was studied in detail by Halpern & Oke (1987). It is located 26.6 arcmin off-axis in a 42.1 ks ROSAT pointing on M31 (we have taken special care

| (1) Name | (2) Mean $L_X$ erg s⁻¹ | (3) Normaliz. at 1 keV | (4) Photon index | (5) Variability description factor of | (6) Fe II/ Hβ | (7) FWHM Hβ km s⁻¹ | (8) Gal $N_H$ ·10²⁰ cm⁻² | (9) Fit $N_H$ | (10) $\chi^2_\nu$ (d.o.f.) |
|---|---|---|---|---|---|---|---|---|---|
| 5C3.100 | $1.3 \cdot 10^{44}$ | $5.6 \cdot 10^{-4}$ | $2.9 \pm 0.2$ | 1.9 in 18900 s | 2.7 | 685 | 6.2 | $9.6 \pm 1.1$ | 0.8 (14) |
| I Zw 1 | $5.0 \cdot 10^{44}$ | $3.4 \cdot 10^{-3}$ | $3.0 \pm 0.1$ | 1.5 in 6200 s | 1.3 | 1240 | 5.0 | $6.5 \pm 0.7$ | 0.4 (10) |
| HB89 0111 − 015 | $3.0 \cdot 10^{43}$ | $1.2 \cdot 10^{-4}$ | $2.3 \pm 1.4$ | — | 0.7 | 1130 | 4.2 | $6.3^{+8.0}_{-6.3}$ | 1.8 (7) |
| Mrk 359 | $2.6 \cdot 10^{43}$ | $3.5 \cdot 10^{-3}$ | $2.4 \pm 0.1$ | 1.5 in 35600 s | 0.08 | 135 | 4.7 | $5.0 \pm 0.7$ | 0.9 (25) |
| PKS 0129 − 066 | $2.5 \cdot 10^{44}$ | $2.3 \cdot 10^{-5}$ | $4.1 \pm 1.8$ | 1.9 in 12000 s (?) | 1.8 | 1310 | 3.7 | $4.0^{+6.3}_{-4.0}$ | 1.3 (12) |
| E0132 − 411 | $5.1 \cdot 10^{44}$ | $2.5 \cdot 10^{-5}$ | $4.1 \pm 0.6$ | 1.3 in 15000 s (?) | 1.2 | 1930 | 2.0 | $3.6 \pm 2.1$ | 0.9 (14) |
| E0144 − 005 | $1.1 \cdot 10^{44}$ | $4.4 \cdot 10^{-4}$ | $2.7 \pm 0.2$ | — | 0.8 | 1940 | 2.8 | $3.3 \pm 0.9$ | 0.5 (21) |
| Mrk 1044 | $6.3 \cdot 10^{43}$ | $5.4 \cdot 10^{-3}$ | $3.0 \pm 0.1$ | 1.4 in 86000 s | 0.9 | 1280 | 3.0 | $4.1 \pm 0.4$ | 1.7 (25) |
| 1E 0919+515 | $4.4 \cdot 10^{44}$ | $1.8 \cdot 10^{-4}$ | $3.7 \pm 0.2$ | 5.9 in $1.5 \cdot 10^{6}$ s | 1.5 | 1390 | 1.4 | $2.8 \pm 0.6$ | 1.3 (20) |
| E0944+464 | $7.6 \cdot 10^{43}$ | $2.6 \cdot 10^{-5}$ | $3.7 \pm 1.4$ | — | 1.3 | 1320 | 1.4 | $3.6 \pm 5.0$ | 1.9 (7) |
| Mrk 1239 | $4.1 \cdot 10^{42}$ | $2.1 \cdot 10^{-4}$ | $3.9 \pm 0.3$ | — | 0.6 | 910 | 3.9 | $8.3 \pm 1.6$ | 1.0 (13) |
| RE J 1034+393 | $3.9 \cdot 10^{44}$ | $1.7 \cdot 10^{-3}$ | $4.3 \pm 0.1$ | 1.1 in 12000 s | — | 1500 | 1.4 | $3.3 \pm 0.2$ | 4.5 (11) |
| Mrk 42 | $1.2 \cdot 10^{43}$ | $4.9 \cdot 10^{-4}$ | $2.6 \pm 0.2$ | 1.3 in 86000 s | 1.0 | 670 | 1.9 | $2.5 \pm 0.7$ | 0.6 (14) |
| Mrk 766 | $9.7 \cdot 10^{43}$ | $5.1 \cdot 10^{-3}$ | $2.5 \pm 0.2^{1}$ | 3 in 40000 s | 0.7 | 850 | 1.7 | $2.1 \pm 0.4^{1}$ | 1.6 (14)[1] |
| 1E 1223.5+2522 | $4.8 \cdot 10^{43}$ | $6.1 \cdot 10^{-5}$ | $3.9 \pm 0.3$ | 1.3 in 294400 s (?) | 0.6 | 1730 | 1.8 | $2.9 \pm 1.0$ | 1.8 (13) |
| 1E 1226.9+1336 | $5.5 \cdot 10^{43}$ | $7.4 \cdot 10^{-5}$ | $2.9 \pm 1.3$ | 2.0 in 80000 s (?) | 1.1 | 1120 | 2.7 | $4.5 \pm 0.6$ | 0.4 (5) |
| E1228+123 | $4.6 \cdot 10^{43}$ | $6.2 \cdot 10^{-5}$ | $3.1 \pm 0.7$ | 2.6 in 100000 s (?) | 1.15 | 1680 | 2.5 | $4.4 \pm 3.2$ | 1.0 (14) |
| RE J 1237+264 | $2.1 \cdot 10^{45}$ | $3.7 \cdot 10^{-3}$ | $4.8 \pm 0.3$ | 1.4 in $1.3 \cdot 10^{7}$ s | 1.4 | 1200 | 1.4 | $5.2 \pm 0.7$ | 1.5 (20) |
| IRAS 13224 − 3809 | $2.8 \cdot 10^{44}$ | $9.2 \cdot 10^{-4}$ | $4.4 \pm 0.2$ | 2.0 in 800 s | 2.4 | 650 | 5.0 | $8.7 \pm 0.5$ | 4.3 (16) |
| E1346+266 | $2.0 \cdot 10^{46}$ | $4.3 \cdot 10^{-5}$ | $4.1 \pm 0.2$ | 2.5 in 133200 s | 0.98 | 1840 | 1.1 | $2.6 \pm 0.5$ | 2.3 (21) |
| Mrk 478 | $6.4 \cdot 10^{44}$ | $1.9 \cdot 10^{-3}$ | $3.6 \pm 0.1$ | — | 0.7 | 1370 | 1.0 | $2.0 \pm 0.3$ | 2.2 (15) |
| Mrk 291 | $8.9 \cdot 10^{42}$ | $2.8 \cdot 10^{-4}$ | $2.1 \pm 0.8$ | 1.4 in 6000 s (?) | 1.1 | <700 | 3.5 | $2.1^{+2.2}_{-2.1}$ | 0.8 (8) |
| Mrk 493 | $3.5 \cdot 10^{43}$ | $7.6 \cdot 10^{-4}$ | $2.7 \pm 0.2$ | 1.3 in $2.2 \cdot 10^{6}$ s | 1.31 | 410 | 2.0 | $2.6 \pm 0.5$ | 1.1 (17) |
| HB89 1557 + 272 | $1.1 \cdot 10^{43}$ | $1.4 \cdot 10^{-4}$ | $1.3 \pm 0.6$ | — | — | 1410 | 3.9 | $2.9 \pm 0.9$ | 0.7 (14) |
| E1640+537 | $3.2 \cdot 10^{43}$ | $6.1 \cdot 10^{-5}$ | $2.1 \pm 1.1$ | — | — | 970 | 2.6 | $2.6^{+3.3}_{-2.6}$ | 0.9 (11) |
| IRAS 1652+395 | $7.9 \cdot 10^{43}$ | $3.0 \cdot 10^{-4}$ | $2.7 \pm 0.2$ | 1.4 in 5800 | 1.3 | 1000 | 1.7 | $2.1 \pm 0.6$ | 1.1 (20) |
| Kaz 163 | $8.2 \cdot 10^{43}$ | $9.4 \cdot 10^{-4}$ | $2.5 \pm 0.1$ | 1.5 in 55000 s | 0.48 | 2110 | 4.3 | $3.7 \pm 0.4$ | 0.9 (29) |
| Mrk 507 | $8.9 \cdot 10^{42}$ | $1.5 \cdot 10^{-4}$ | $1.6 \pm 0.3$ | 1.8 in 17720 s | 2.7 | 960 | 4.3 | $3.8 \pm 0.6$ | 1.1 (16) |
| E1805+700 | $1.0 \cdot 10^{44}$ | $8.3 \cdot 10^{-5}$ | $2.5 \pm 0.5$ | — | 1.2 | 1700 | 4.8 | $4.4 \pm 2.4$ | 0.6 (9) |
| Mrk 896 | $4.5 \cdot 10^{43}$ | $1.4 \cdot 10^{-3}$ | $2.6 \pm 0.1$ | 1.9 in 293000 s | — | 1330 | 4.7 | $3.4 \pm 0.4$ | 1.5 (24) |
| Ark 564 | $2.3 \cdot 10^{44}$ | $1.2 \cdot 10^{-2}$ | $3.4 \pm 0.1$ | 1.3 in 1600 s | 0.8 | 720 | 6.1 | $7.3 \pm 0.4$ | 2.2 (29) |
| E0337 − 267 | — | — | — | — | 1.1 | 1340 | 0.9 | — | — |
| 1E 0440.0 − 1057 | — | — | — | — | 0.85 | 2010 | 5.4 | — | — |
| H0707 − 495 | — | — | — | — | $2.77^{2}$ | 1000 | 4.4 | — | — |
| PG 1016+336 | — | — | — | — | — | 1600 | 1.6 | — | — |
| E1028+310 | — | — | — | — | <0.9 | 1780 | 1.9 | — | — |
| 1E 1031.1+5822 | — | — | — | — | — | 760 | 0.6 | — | — |
| 1E 1205.5+4657 | — | — | — | — | — | 1470 | 1.5 | — | — |
| NGC 4748 | — | — | — | — | — | 1470 | 3.7 | — | — |
| Mrk 783 | — | — | — | — | <0.11 | 1900 | 2.0 | — | — |
| Mrk 684 | — | — | — | — | — | 1400 | 1.5 | — | — |
| PG 1448+273 | — | — | $3.2 \pm 0.3^{3}$ | — | — | 820 | 2.4 | — | — |
| IRAS 15091 − 2107 | — | — | — | — | — | 1480 | 8.3 | — | — |
| E1511+671 | — | — | — | — | 0.89 | 1650 | 2.4 | — | — |
| UCM 2257+2438 | — | — | — | — | — | 1100 | 4.9 | — | — |
| Mrk 1126 | — | — | — | — | <0.01 | 2500 | 3.5 | — | — |

**Table 1.** Properties of NLS1. The first 31 NLS1 in column 1 are those for which we present ROSAT data, and they are listed in right ascension order. The rest of the NLS1 in column 1 are those for which there is no public ROSAT data currently, and they are listed separately in right ascension order. We present mean 0.1–2.4 keV isotropic luminosities derived using the power-law model (column 2), the ROSAT power-law fit normalization in units of photons cm⁻² s⁻¹ keV⁻¹ at 1 keV (column 3), the ROSAT power-law fit photon index (column 4), a variability description (column 5), the Fe II 4570 Å/Hβ line flux ratio (column 6), the full width at half maximum of Hβ (column 7), the Galactic $N_H$ value from Stark et al. (1992) (column 8), the ROSAT power-law fit $N_H$ value (column 9) and the reduced $\chi^2$ and number of degrees of freedom of the ROSAT power-law fit (column 10). Confidence intervals for the ROSAT fit parameters are 90 per cent intervals. The question marks in column 5 indicate that the variability given for the objects is marginal; cf. Sects. 4.7 and 4.10 for details. [1] The values are taken for the very low state as defined by Molendi & Maccacaro (1994). [2] The Fe II line intensity given is summed over the multiplets 4570, 5190 and 5320 Å. [3] This photon index is taken from Walter & Fink (1993) and its error is for 68.3 per cent confidence.

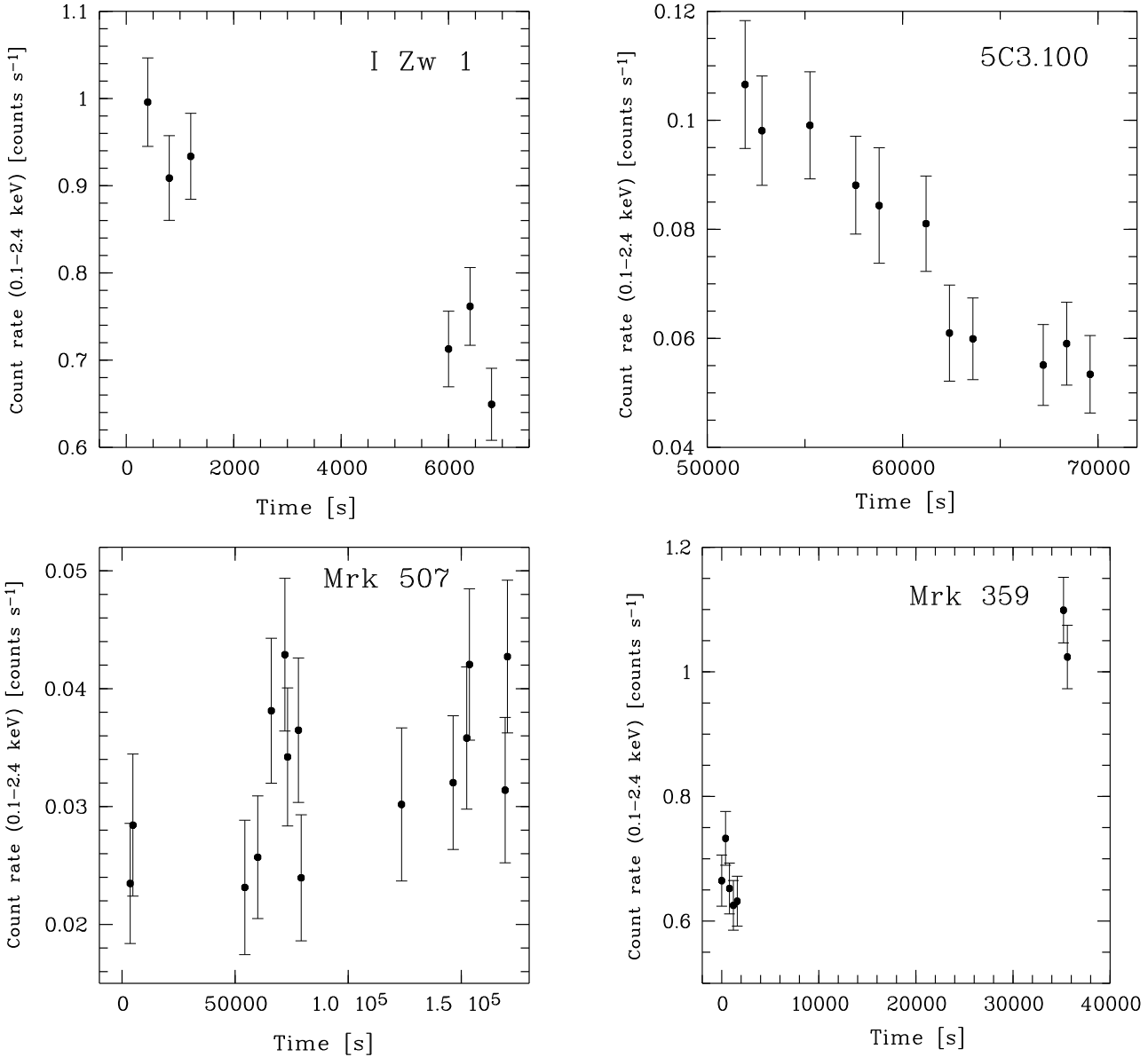

**Fig. 3.** X-ray light curves of I Zw 1, 5C3.100, Mrk 507 and Mrk 359. The bin size is 400 seconds for I Zw 1 and Mrk 359 and 1200 seconds for 5C3.100 and Mrk 507.

to be sure that the PSPC window support structure does not contaminate its variability). Variability and spectral properties are summarized in Table 1 and Fig. 3. The $N_H$ value obtained from the fit is somewhat larger than the Galactic value of $6.2 \cdot 10^{20}$ cm$^{-2}$, probably due to intrinsic absorption as well as absorption within the disc of M31 which lies along the line of sight.

We have divided the ROSAT observation interval into two subsets depending on whether the count rate is below or above 0.075 count s$^{-1}$ and have computed hardness ratios for each subset. We find indications for an anticorrelation between hardness ratio and source intensity. For the low ($< 0.075$ count s$^{-1}$) and the high ($> 0.075$ count s$^{-1}$) states the calculated hardness ratios are $0.1 \pm 0.03$ and $0.0 \pm 0.05$.

### 4.4. Mrk 507 (IRAS 17489+6843)

Since optical spectroscopy of the luminous IRAS AGN Mrk 507 (Halpern & Oke 1987) revealed that the object is a strong optical Fe II emitter, we proposed the object for a ROSAT AO-4 PSPC pointing. X-ray flux variations were detected (see Fig. 3). No spectral variations are detected when the observation is divided into two subsets ranging from 0.02–0.03 count s$^{-1}$ and 0.03–0.05 count s$^{-1}$. The hardness ratio for the low state is $-0.41 \pm 0.04$ and the hardness ratio for the high state is $-0.44 \pm 0.03$. We note that the spectrum of this source is reasonably well constrained and appears remarkably flat (cf. Table 1). Mrk 507 has a tortuous history of classification and re-classification as is described in Halpern & Oke (1987). At times it has been called a LINER, Seyfert 1, Seyfert 2, LINER/Seyfert 2 transition object, narrow-line Seyfert 1, or been denied classification altogether due to its peculiar properties. Our X-ray observations are not able to settle its classification, but we note that it does not follow the general trend of steep soft X-ray spectra seen in other NLS1 (cf. Sect. 5.1). It may be an exceptional object.

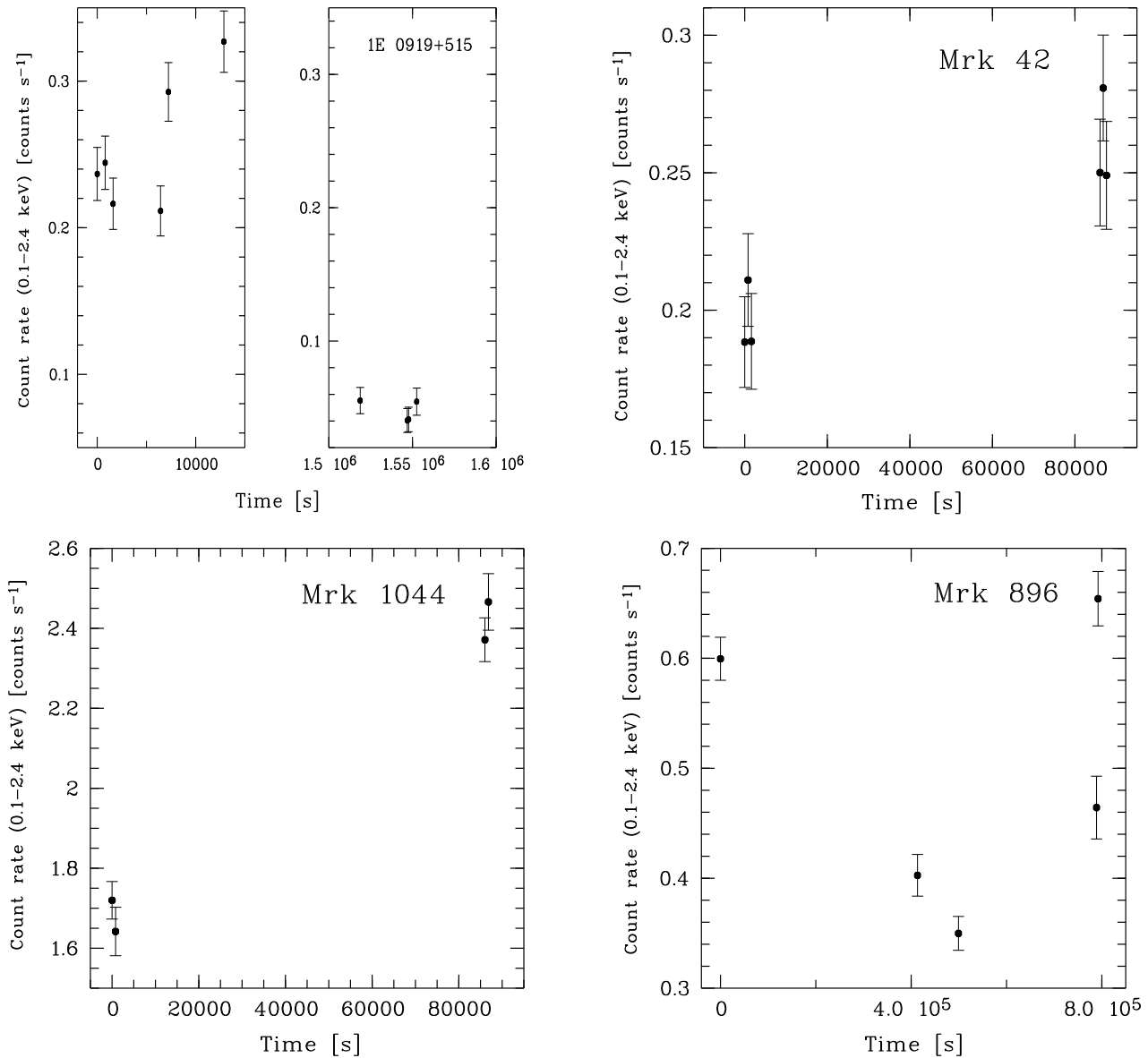

**Fig. 4.** X-ray light curves of 1E 0919+515, Mrk 42, Mrk 1044 and Mrk 896. The bin size is 1600 seconds for Mrk 896 and 800 seconds for the other objects.

### 4.5. Ark 564 (MCG+05−53−012, UGC 12163)

Ark 564 is a bright NLS1. It has relatively strong [Ca II] emission and strong Fe II emission lines (van Groningen 1993). Ark 564 was one of the peculiar outlying Seyferts in the Walter & Fink (1993) ROSAT all-sky survey correlation between ROSAT spectral slope and ultraviolet (1375 Å) to 2 keV flux ratio and has been the subject of a recent detailed X-ray study by Brandt et al. (1994). Spectral analysis of Ark 564 consistently yields very steep ($\Gamma \approx 3$ where $\Gamma$ is the photon spectral index) spectra. A peculiar edge-like feature is also seen at $1.15^{+0.06}_{-0.05}$ keV. In the absence of flattening at energies below 0.1 keV, the extrapolated soft X-ray spectrum overpredicts the dereddened 1375 Å and 2675 Å fluxes (Walter & Fink 1993) by factors > 160 and > 120, respectively. Ark 564's 0.1–2.4 keV flux varies by ∼ 20 per cent in 1500 s, and further long term variability by a factor of about three is seen in archival HEAO 1 data (Brandt et al. 1994).

### 4.6. Mrk 766 (NGC 4253)

The flux and spectral variability of Mrk 766 were studied by Molendi et al. (1993) and Molendi & Maccacaro (1994) based both on ROSAT all-sky survey and pointed observations. Mrk

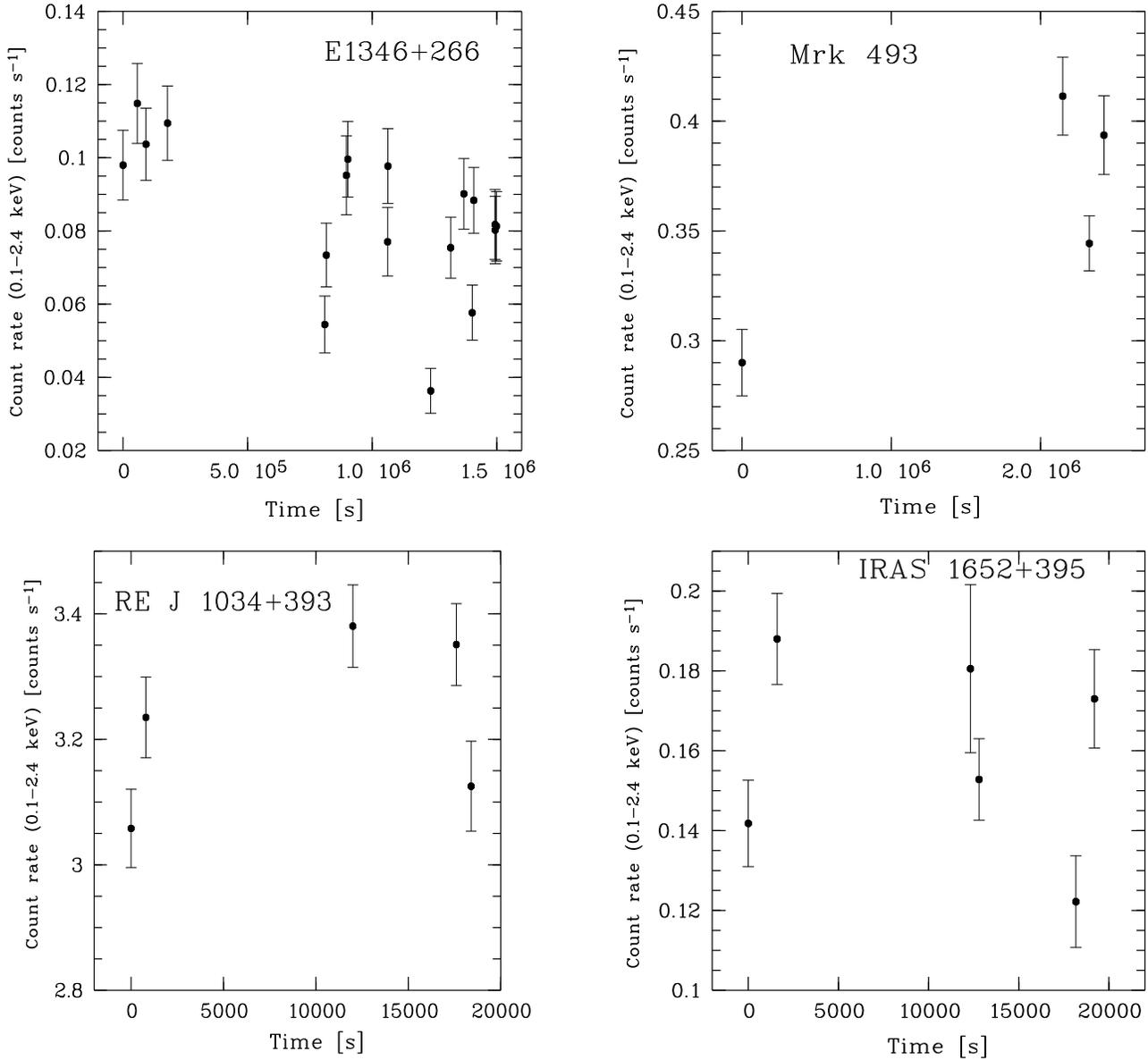

**Fig. 5.** X-ray light curves for E1346+266, Mrk 493, RE J 1034+393 and IRAS 1652+395. The bin size is $1.2 \cdot 10^5$ seconds for Mrk 493, 1600 seconds for IRAS 1652+395, 1200 seconds for E1346+26 and 800 seconds for RE J 1034+393.

766 is found to be variable by a factor of about 3 on timescales of a few hours. Molendi & Maccacaro (1994) have found spectral variability in which the 0.1–0.9 keV part of the spectrum hardens as the source brightens while the 0.9–2.0 keV part of the spectrum does not change significantly. They claim that these variations cannot be explained by changes in the optical depth of an absorption edge or changes in the ionization parameter and/or column density of a warm absorber. Emission from an accretion disc, however, can explain the observed spectrum and spectral variations. Mrk 766 was another of the outlying Seyferts in the Walter & Fink (1993) ROSAT all-sky survey correlation between spectral slope and ultraviolet (1375 Å) to 2 keV flux ratio.

*4.7. PKS 0129−066 (E0129−066), E0132−411, RE J 1034+393 (KUG 1031+398), RE J 1237+264 (IC 3599, ZW 159.034) and E1346+266*

These objects all have extremely steep soft X-ray continua with best fit power-law photon indices *greater than or equal to 4.* The first two were listed as ultrasoft AGN in Puchnarewicz et al. (1992). E0132−411 is presented in detail in Sect. 6 of Puchnarewicz et al. (1992) and also in Thompson et al. (1994).

The third and fourth are two of the three ultrasoft Seyferts (all are NLS1) from the ROSAT Wide Field Camera Bright Source Catalogue (Pounds et al. 1993; Shara et al. 1993). The best power-law fit to the X-ray spectrum of RE J 1034+393 has a photon index of $4.3 \pm 0.1$, but a power-law model is a poor description of the ROSAT spectrum. More detailed spec-tral analyses may be found in Pounds (1994), Pounds & Done (1995) and Puchnarewicz et al. (1995). RE J 1237+264 is located in several ROSAT pointings and has been modelled by Brandt, Pounds & Fink (1995) using both all-sky survey and pointed data. They find an ultrasoft spectrum with an all-sky survey photon index of $4.8 \pm 0.3$. In addition, in the year between the all-sky survey and pointed observations the 0.1–2.4 keV ROSAT count rate dropped by a factor of $\sim 70$ without evidence for spectral change. Brandt, Pounds & Fink (1995) present a positional analysis and evidence against contaminating sources.

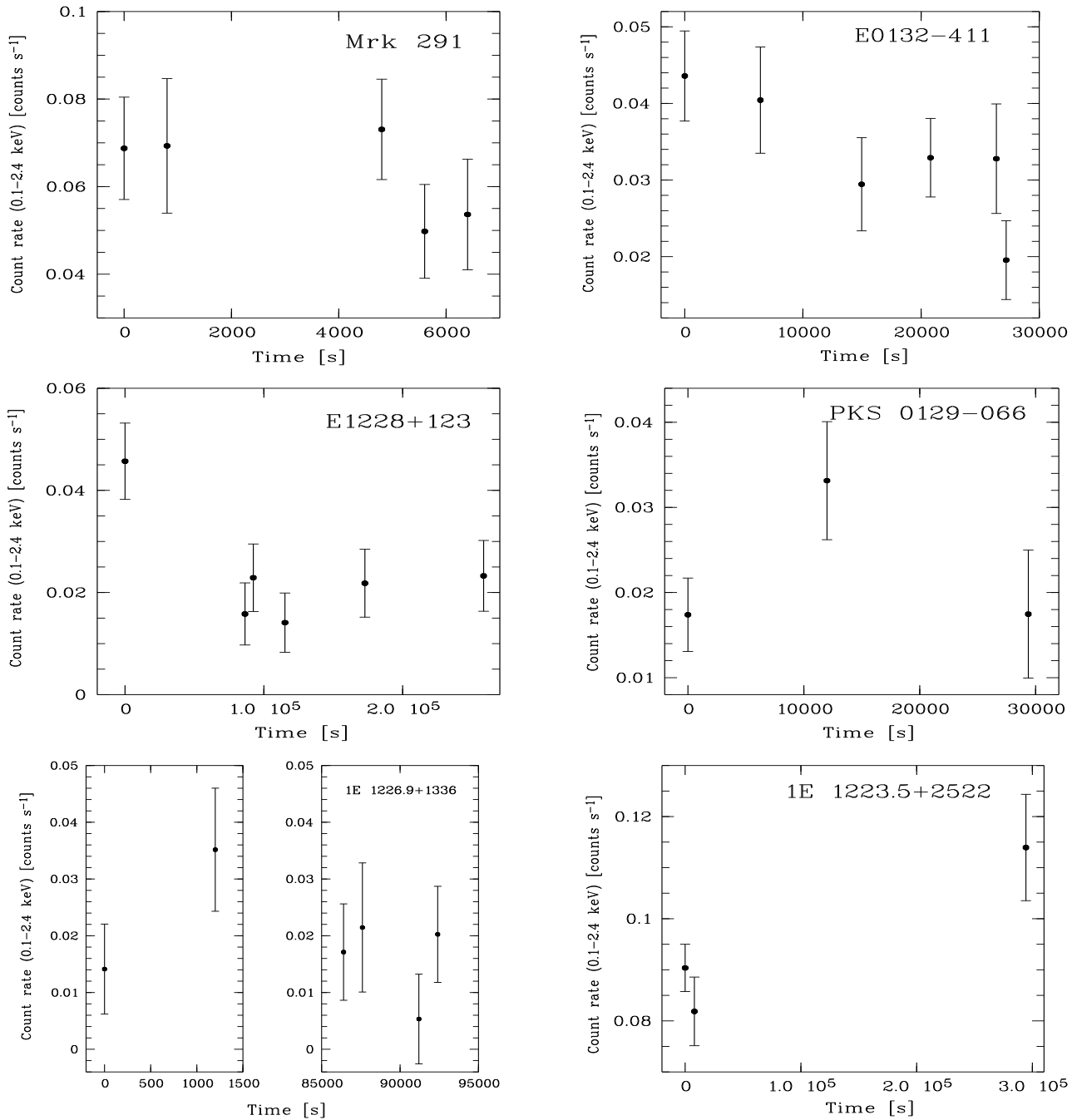

**Fig. 6.** X-ray light curves for Mrk 291, E0132 − 411, E1228+123, PKS 0129 − 066, 1E 1226.9+1336 and 1E 1223.5+2522. The bin size is 8000 seconds for 1E 1223.5+2522, 2000 seconds for PKS 0129 − 066, 1600 seconds for E0132 − 411, 1200 seconds for E1228+123 and 1E 1226.9+1336 and 800 seconds for Mrk 291. These objects show marginal evidence for variability.

E1346+266 is a narrow line quasar at a redshift of 0.92. Puchnarewicz et al. (1994) discuss the remarkably high temperature soft X-ray excess implied by E1346+266's redshift and ROSAT spectrum, and they estimate its entire big blue bump isotropic luminosity to be $\sim 10^{47}$ erg s$^{-1}$. E1346+266 is located about 10 arcmin off-axis in a 36 ks pointing on Abell 1795. It appears to be an extension of the NLS1 class to very high luminosity.

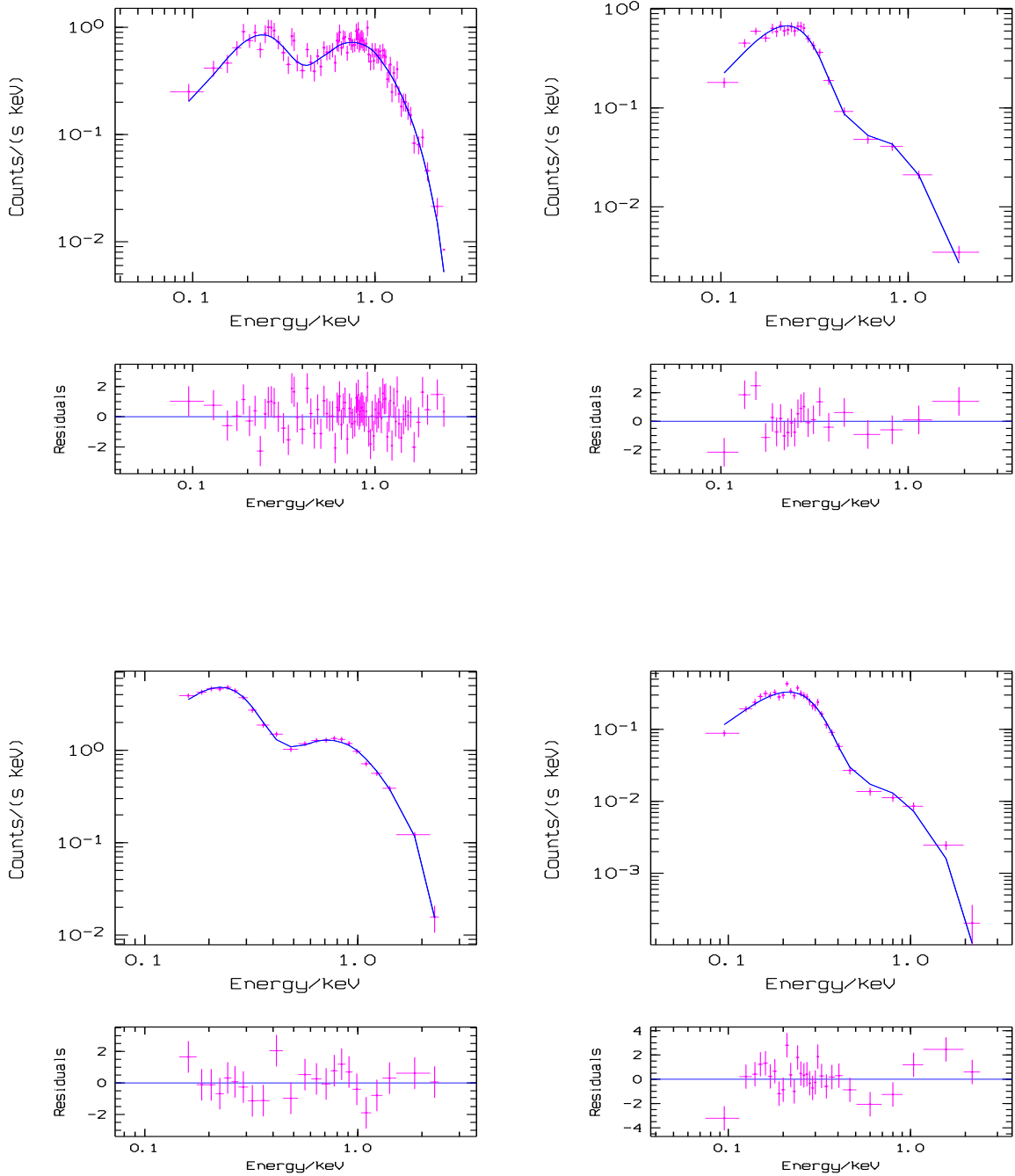

**Fig. 7.** Soft X-ray spectra of I Zw 1 (left upper corner), Mrk 1044 (left lower), 1E 0919+515 (right upper) and E1346+266 (right lower). Simple power law fits indicate extremely steep X-ray continua with photon indices of $3.0 \pm 0.1$ (I Zw 1), $3.0 \pm 0.1$ (Mrk 1044), $3.7 \pm 0.2$ (1E 0919+515) and $4.1 \pm 0.2$ (E1346+266). Although a simple power-law fit does not always result in a satisfactory description of the spectrum, the photon index serves as a rough quantitative measure of its steepness. References to more detailed spectral analyses for some objects can be found in Sect. 4.

| (1) Target name | (2) Observation date | (3) Exp. (ks) | (4) Source counts | (5) z |
|---|---|---|---|---|
| 5C3.100 | Jul 14–15 1991 | 42.1 | 3170 | 0.071 |
| I Zw 1 | Dec 31 1991 | 3.5 | 2780 | 0.060 |
| HB89 0111 – 015 | Jul 13–13 1992 | 5.7 | 100 | 0.12 |
| Mrk 359 | Jul 15–15 1992 | 3.2 | 2518 | 0.017 |
| PKS 0129 – 066 | Jul 12–12 1992 | 4.1 | 85 | 0.22 |
| E0132 – 411 | Jul 6–7 1992 | 9.4 | 303 | 0.27 |
| E0144 – 005 | Jul 19–20 1992 | 6.0 | 292 | 0.08 |
| Mrk 1044 | Aug 7–10 1992 | 2.8 | 5850 | 0.016 |
| 1E 0919+515 | Feb 10–13 1992 | 13.4 | 2422 | 0.16 |
| E0944+464 | Apr 14–18 1992 | 2.6 | 66 | 0.35 |
| Mrk 1239 | Nov 6–8 1992 | 9.1 | 572 | 0.019 |
| RE J 1034+393 | Nov 18–19 1992 | 4.7 | 15400 | 0.042 |
| Mrk 42 | Dec 2–5 1992 | 4.4 | 1072 | 0.024 |
| Mrk 766 | Jun 16–21 1992 | 6.8 | 24000 | 0.013 |
| 1E 1223.5+2522 | May 31–Jun 5 1992 | 8.4 | 770 | 0.24 |
| 1E 1226.9+1336 | Dec 24–26 1991 | 7.4 | 133 | 0.15 |
| E1228+123 | Jun 10–13 1992 | 10.5 | 265 | 0.12 |
| RE J 1237+264 | Jun 18 1992 | 2.4 | 96 | 0.021 |
| IRAS 13224 – 3809 | Aug 10–12 1992 | 20.1 | 6200 | 0.067 |
| E1346+266 | Jan 2–22 1992 | 36.8 | 2750 | 0.92 |
| Mrk 478 | Jan 17–17 1992 | 2.4 | 5046 | 0.079 |
| Mrk 291 | Feb 27–27 1992 | 3.8 | 263 | 0.035 |
| Mrk 493 | Jan 30–Feb 29 1992 | 8.2 | 2903 | 0.032 |
| HB89 1557+272 | Jul 20–21 1992 | 7.7 | 225 | 0.065 |
| E1640+537 | Jan 1–3 1992 | 5.0 | 97 | 0.14 |
| IRAS 1652+395 | Feb 24–26 1991 | 7.6 | 1197 | 0.069 |
| Kaz 163 | Aug 8–9 1993 | 24.8 | 6710 | 0.063 |
| Mrk 507 | Aug 8–10 1993 | 24.7 | 790 | 0.056 |
| E1805+700 | May 13–17 1992 | 10.5 | 203 | 0.19 |
| Mrk 896 | Nov 9–10 1992 | 6.9 | 3660 | 0.026 |
| Ark 564 | Nov 30 1993 | 4.3 | 13080 | 0.024 |

**Table 2.** Observation log for NLS1 ordered by right ascension: target name (column 1); ROSAT observation date (column 2); exposure time in ks (column 3); total source counts after background subtraction and correction for vignetting (column 4); redshift (column 5).

When detailed spectral modelling is performed for RE J 1034+393, RE J 1237+264 and E1346+266 they all appear to have hot soft excesses with rest-frame temperatures $\sim 10^6$ K. In RE J 1034+393 and RE J 1237+264 the soft excesses show evidence for turning over within the ROSAT band.

### 4.8. Kaz 163 (1E 1747.3+6836, VII Zw 742)

Kaz 163 is a NLS1 in the process of collision with an object to its north resembling an elliptical galaxy in shape and colour (Kriss & Canizares 1982; Hutchings & Hickson 1988). It lies in an irregular blue nebulosity and its nuclear spectrum has relatively weak [O III] lines and some Fe II blends (Hutchings & Hickson 1988). The ratio of its permitted and forbidden lines falls away from the nucleus, but then rises again in outer parts of the nebulosity. Kaz 163 was observed in the *Einstein* EMSS (Kriss & Canizares 1982; Stocke et al. 1991) and the ROSAT all-sky survey (Walter & Fink 1993) but no spectral features or variability were observed. Brandt et al. (1994) have recently discovered rapid X-ray variability (cf. their Fig. 2) as well as evidence for soft spectral complexity.

### 4.9. Mrk 359, 1E 0919+515, Mrk 1044 (MCG–2–7–24), Mrk 42, Mrk 896 and Mrk 493

These objects show some evidence for X-ray variability as well, but with a comparatively slow rate of change (cf. Table 1 and

Figs. 3–5). Mrk 359 has the smallest broad-line widths of any known Seyfert 1 and its narrow lines are also among the narrowest (Veilleux 1991). Veilleux (1991) argues that the abnormally small width of the broad H$\alpha$ line is difficult to explain via aspect effects alone and that there is probably a deficit of high velocity line emitting gas in the center of Mrk 359. It was detected in X-rays during the Einstein slew survey (Elvis et al. 1992) and it is unusual among the NLS1 in our sample since it has both a relatively flat spectral shape and a relatively small Fe II/H$\beta$ ratio. 1E 0919+515 shows X-ray flux variations by a factor of five on a time scale of about $1.5 \cdot 10^6$ s. Mrk 1044 and Mrk 42 have been classified as NLS1 in Osterbrock & Pogge (1985).

### 4.10. Mrk 291 (IRAS 15529+1920), 1E 1226.9+1336, E1228+123 and 1E 1223.5+2522

These sources show marginal evidence for variability, but the variability cannot be proven with high statistical significance. Mrk 291 was classified as a NLS1 galaxy in Goodrich (1989). Its spectrum does not appear to be especially steep but is poorly constrained due to its faintness. 1E 1226.9+1336 and E1228+123 are listed in the Lípari et al. (1993) sample of ultraluminous IRAS galaxies with extreme optical Fe II emission. 1E 1223.5+2522 is discussed in Stephens (1989) and is quite steep in the ROSAT band.

### 4.11. HB89 0111–015, E0144–005, E0944+464, Mrk 1239, Mrk 478 (PG 1440+356), HB89 1557+272 (1E 1557+272), E1640+537 (87GB 164053.0+534543), IRAS 1652+395 and E1805+700

These objects show no indications for variability in their X-ray light curves obtained during the ROSAT pointed observations. Mrk 1239 has been defined as a NLS1 in Osterbrock & Pogge (1985). Mrk 478 is one of the three ultrasoft Seyferts detected with the ROSAT Wide Field Camera. Gondhalekar et al. (1994) have presented a detailed study of Mrk 478. Its ultraviolet/soft X-ray continuum energy distribution is generally consistent with that expected from a thin accretion disc, although there is some discrepancy at energies above about 0.4 keV. The source E1805+700 is located about 19 arcmin off-axis in the ROSAT PSPC field of view. The source is partially hidden by the PSPC support grid, which makes the variability analysis difficult. E1640+537 and E1805+700 were classified as probable ultrasoft sources based on Einstein data (Puchnarewicz et al. 1992).

## 5. Results on NLS1 as a class

### 5.1. Steep soft X-ray spectrum AGNs

Most AGN are known to have relatively flat 2–10 keV X-ray continua with underlying photon indices around 1.9–2.0. Many also have soft excesses below about 0.5–1 keV where the effective photon index rises to about 2.5.

As a class NLS1 generally appear to have steeper soft X-ray spectra than normal Seyfert 1s. In Fig. 8 we illustrate this by showing a plot of ROSAT 0.1–2.4 keV photon index versus FWHM of the H$\beta$ line. We plot NLS1 as filled dots and normal Seyfert 1s as vertical bars. The normal Seyfert 1s in Fig. 8 are those from the systematic ROSAT all-sky survey study of Seyfert 1 galaxies of Walter & Fink (1993) that have H$\beta$

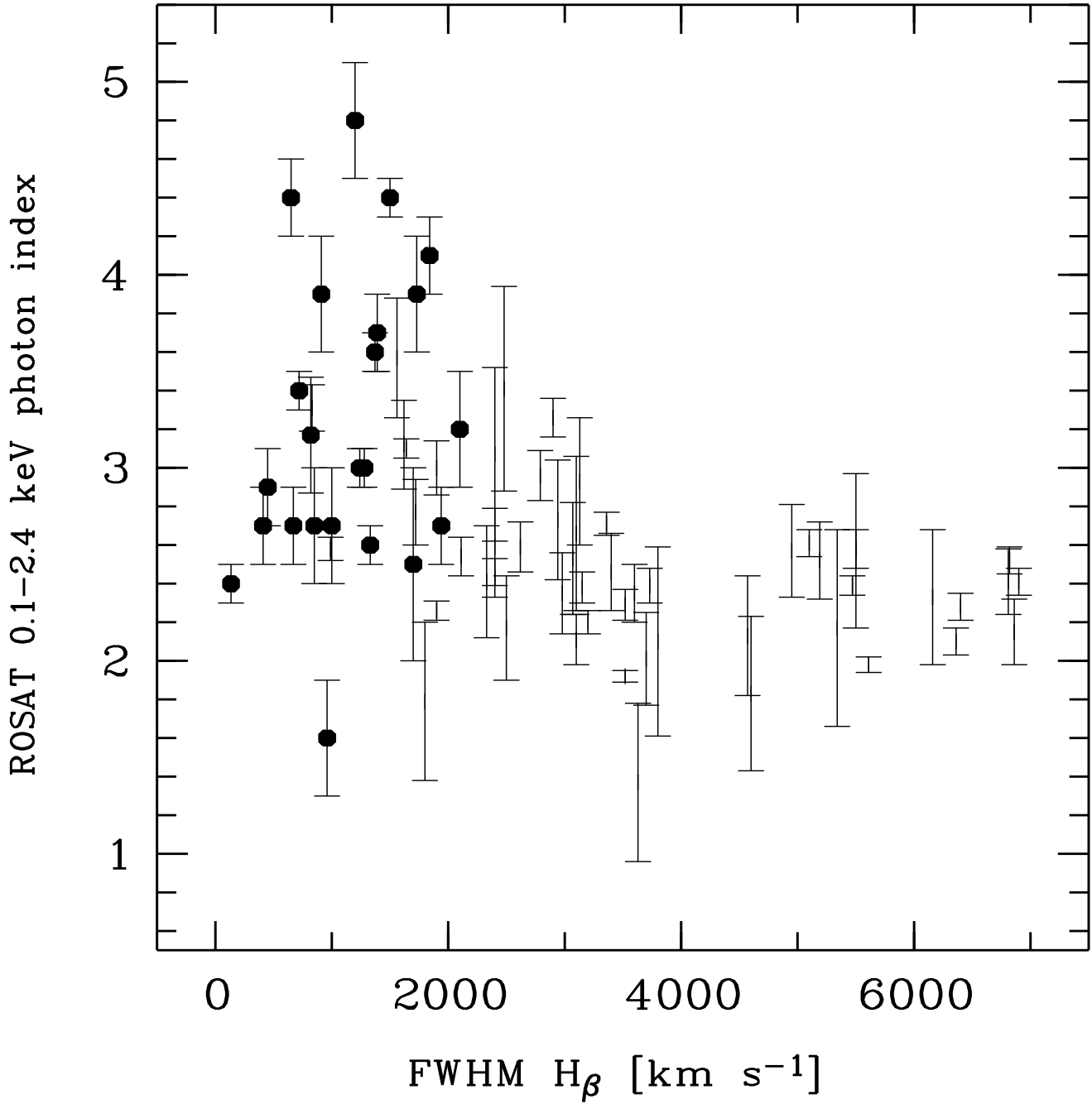

**Fig. 8.** X-ray continuum slope, obtained from a power-law fit, versus FWHM of Hβ for Seyfert 1 galaxies (vertical bars; see the text) and NLS1 (filled dots). Only NLS1 with photon index errors smaller than or equal to 0.5 are plotted. Note that the potential NLS1 with a small photon index is Mrk 507, which may be an exceptional object (cf. Sect. 4.4).

FWHM values in the literature. Additionally, we have used simple power-law fit photon indices derived from pointed data in order to reduce error bar sizes for some normal Seyfert 1s in the Walter & Fink (1993) sample. [1] We have also added some other objects in the literature that have been targets of ROSAT pointed observations. [2] If we consider the 58 Seyfert 1 type galaxies studied by Walter & Fink (1993) and remove the seven NLS1 in their sample, the weighted mean photon index of the resultant sample is 2.34 and the uncertainty in this mean is 0.03 (we compute the uncertainty in the mean using the 68.3 per cent confidence intervals derived from the Walter & Fink 1993 spectral fits and equation 5-15 of Bevington 1969). This is in contrast to our sample of NLS1 which has a weighted mean photon index of 3.13 with an uncertainty in the mean of less than 0.03 (we compute this upper limit conservatively using the 90 per cent confidence intervals from Table 1 and equation 5-15 of Bevington 1969). We note that these general spectral results are very robust to possible changes in the ROSAT response matrix due to the fact that ROSAT observed NLS1 are steep *even when compared to* ROSAT observed normal Seyfert 1s.

One limitation of Fig. 8 that must be kept in mind is that simultaneous ROSAT observations and optical measurements of H$\beta$ FWHM are not available. We are aware that line profiles in Seyferts can change with time (cf. Penston & Pérez 1984; Iijima, Rafanelli & Bianchini 1992 and Loska, Czerny & Szczerba 1993 for dramatic examples of line changes in 3C390.3, NGC 4151 and NGC 5548), but due to the relatively large number of objects in our sample we suspect that optical line changes do not affect the essential nature of our results.

In Fig. 8 the NLS1 and normal Seyfert 1 groups appear to merge continuously with each other. This is in harmony with the work of Goodrich (1989), who argued that the 'differences' between NLS1 and normal Seyfert 1s really represent a continuum of properties, and we will return to this point in our discussion below. Due to the fact that some objects with small H$\beta$ FWHM and small photon indices appear to exist, it is not entirely clear whether the distribution of objects in Fig. 8 forms an anticorrelation between ROSAT continuum slope and H$\beta$ FWHM (as is discussed in Laor et al. 1994) or an exclusion of large H$\beta$ FWHM objects that have steep ROSAT slopes as well (cf. several of the flatter objects in Table 1 that were not plotted in Fig. 8 due to their larger error bars).

Fig. 8 is one of our main results and raises two fundamental questions that we shall address in Sect. 6:

1. Why are Seyfert 1s with $\Gamma > 3$ as well as FWHM H$\beta >$ 3000 km s$^{-1}$ discriminated against by nature when Seyfert 1s can have either of these properties separately?
2. What is the underlying physics that decides the position of a Seyfert 1 along the distribution shown in Fig. 8?

In the NLS1 where detailed spectral modelling has been performed, it often but not always appears that the spectrum can be fit with a hard power law and a soft excess component. Sometimes soft excess components are considered to arise when flux from the hard power law is reflected and reprocessed by matter near the central engine (e.g. Brandt et al. 1993; Czerny & Życki 1994). This interpretation runs into difficulty with at least some NLS1 due to the fact that, barring any dramatic high energy behaviour and assuming that there is no strong obscuration which blocks the hard X-rays yet not the soft X-rays, *the soft excess contains more luminosity than the hard tail* (e.g. Brandt, Pounds & Fink 1995). In at least some NLS1 it therefore appears that we are seeing mostly *intrinsic* emission by matter, perhaps in an accretion disc or optically thin plasma, near the central black hole.

The steep slopes of NLS1 in the ROSAT band might be formed in either of two ways (or via a combination of the two):

1. NLS1 have stronger than usual soft X-ray excess components relative to their near IR and optical spectra and have relatively normal hard X-ray tail components.
2. NLS1 have normal soft X-ray excess components relative to their near IR and optical spectra and have weaker than usual hard X-ray tail components.

We shall examine this issue systematically in future work on the broad band spectral energy distributions of NLS1, but current evidence suggests that for some NLS1 (E1346+266, RE J1034+393, Mrk 478) the first case applies (Sect. 8.3.1 of Puchnarewicz et al. 1992; Sect. 5 of Puchnarewicz et al. 1995; Sect. 3 of Fiore & Elvis 1995) while for others the second case applies (cf. Laor et al. 1994). We shall therefore consider both possibilities in our discussion below.

The low-luminosity and well-studied Seyfert NGC 4051 lies among the NLS1 in Fig. 8, and it shares the rapid X-ray variability of many of them (e.g. Lawrence et al. 1987; McHardy et al. 1995). NGC 4051 has a H$\beta$ FWHM $\approx$ 990 km s$^{-1}$ (Osterbrock & Shuder 1982), strong Fe II lines (Lípari et al. 1993) and a ratio of [O III] 5007 Å to H$\beta$ of $\approx$ 1.0 (Osterbrock & Shuder 1982). It therefore satisfies the descriptive criteria of Goodrich (1989) for NLS1 and might be fruitfully thought of as a NLS1 (as we discuss in detail below we work in the context that NLS1 and normal Seyfert 1s are essentially the same types of objects and thus do not wish to be pedantic about taxonomy). Pounds et al. (1994) claim that NGC 4051 is a bare nucleus Seyfert 1 galaxy seen nearly pole-on (cf. our discussion of the pole-on model for NLS1 below) while Bao & Ostgaard (1994) claim to detect a drop-off in its X-ray variability power spectrum which indicates a relatively small black hole mass of $\sim 4 \cdot 10^6$ solar masses (cf. our discussion of NLS1 black hole masses below; Sect. 6.2 of Papadakis & Lawrence 1995 gives an upper limit of $7 \cdot 10^6$ solar masses). NGC 4051 is one of the peculiar outlying Seyferts in Fig. 8 of Walter & Fink (1993) and they state it has a peculiarly hot soft X-ray excess (cf. Mihara et al. 1994). It is also one of the three 'non-typical' Seyfert 1s of White, Fabian & Mushotzky (1984). Marshall et al. (1983) were the first to note its unusually soft X-ray spectrum. In addition, a recent ASCA observation reported by Mihara et al. (1994) has found complex absorption by warm matter with edges at approximate energies of 0.74, 0.90 and 1.4 keV. Mrk 142, Mrk 335, PG 1244+026 and PG 1543+489 also lie among the NLS1, as they all have small H$\beta$ FWHM and steep ROSAT slopes. They satisfy the other requirements of Goodrich (1989) for classification as NLS1 (cf. Lee et al. 1988; Lípari, Terlevich & Macchetto 1993; Laor et al. 1994). Lee et al. (1988) have reported a soft X-ray flare in Mrk 335, in line with the rapid variability we have described in many of these objects.

Another object worthy of mention in connection with NLS1 is PKS 0558−504 (Remillard et al. 1986). Unfortunately there are no pointed ROSAT PSPC observations of this object. PKS 0558 − 504 has a H$\beta$ FWHM of 1500 km s$^{-1}$ and satisfies the other Goodrich (1989) NLS1 criteria as well. Remillard et al. (1991) discovered a rapid energetic X-ray flare in PKS 0558 − 504 which demands that its apparent luminosity was enhanced by relativistic beaming. H0707−495 (Remillard et al. 1986) also satisfies the Goodrich (1989) NLS1 criteria but again lacks a pointed ROSAT PSPC observation.

and Mrk 335 (Turner, George & Mushotzky 1993).

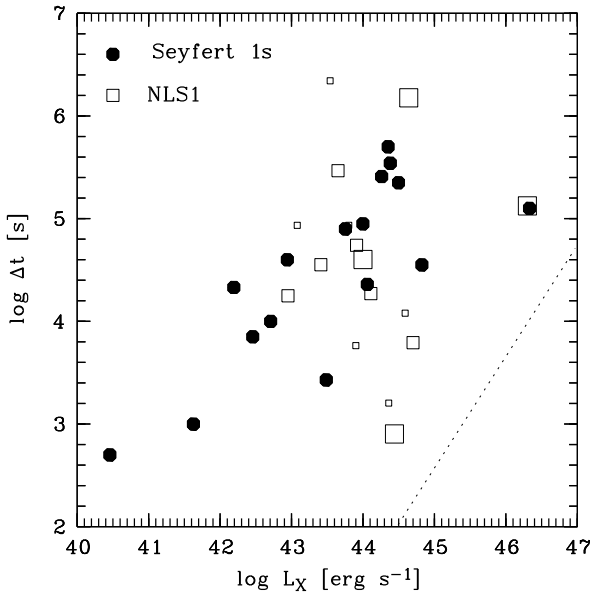

**Fig. 9.** Variation time $\Delta t$ versus X-ray isotropic luminosity of NLS1 (open rectangles) and normal Seyfert 1 galaxies (filled hexagons). The normal Seyfert 1 galaxies are taken from Wandel & Mushotzky (1986). NGC 6814 has not been included since its variability is probably contaminated (Madejski et al. 1993). The size of the NLS1 rectangles corresponds to the maximum variation detected in the ROSAT light curves of the objects: above or equal to a factor of 2, large; factor 1.5–2, medium; and factor below 1.5, small. The dashed line gives the Eddington luminosity for an upper limit on the black hole mass resulting from the time variability constraint. We assume that the X-ray emitting region is located at a distance $R \leq 5R_S$, where $R_S$ is the Schwarzschild radius.

### 5.2. The soft X-ray variability of NLS1

Many NLS1 show rapid soft X-ray variability (cf. Table 1). In Fig. 9 we compare the variation times of NLS1 in our sample with normal Seyfert 1 galaxies taken from Wandel and Mushotzky (1986). The 2–10 keV luminosities of Wandel and Mushotzky (1986) have been extrapolated to the ROSAT band assuming a broken power law model with $\Gamma_1 = 1.9$ above 2 keV and $\Gamma_2 = 2.5$ below 2 keV. The luminosities are corrected for the Galactic foreground value of $N_H$ at the source position.

Two important conclusions may be drawn from rapid soft X-ray variability:

1. Rapid large amplitude variations indicate that substantial fractions of the soft X-rays from NLS1 come from compact regions less than a light day in size. This is smaller than the characteristic size of typical Seyfert 1 BLRs as determined by reverberation mapping observations, demonstrating that NLS1 soft excesses do not originate from extended soft emission by BLR gas or hard X-ray reprocessing by BLR clouds.
2. Variability on short timescales argues that most of the soft X-ray emission from NLS1 must be seen directly from the emitting regions without significant electron scattering. Electron scattering smears out variability and the warm scattering mirrors of Seyfert 2s are thought to have sizes of $\sim 50$ pc based on photoionization modelling (Miller, Goodrich & Matthews 1991) and Hubble Space Tele-scope multiaperture spectropolarimetry (Antonucci, Hurt & Miller 1994). It also worth noting the very large intrinsic soft X-ray luminosities NLS1 would be required to have if their X-rays were electron scattered to us by a Thomson thin mirror.

In addition, IRAS 13224 − 3809 and RE J 1237+264 have shown *giant* amplitude soft X-ray variability (cf. their descriptions in Sect. 4). It is worth noting that NGC 1068, which Ulvestad, Antonucci & Goodrich (1995) suggest would be a NLS1 galaxy if it were viewed along the symmetry axis of its nucleus, may have also shown a spectacular optical (and perhaps X-ray) flare in 1890 (de Vaucouleurs 1991).

The substantial soft X-ray variability and steep slopes of NLS1 echo the relationship discussed in Sect. 3.4.2 of Green, McHardy & Lehto (1993).

### 5.3. Soft X-ray absorption by cold matter

X-ray spectral fits to our sample of NLS1 generally do not show evidence for large amounts of cold absorption over the Galactic column. The weighted mean difference between the value of the neutral hydrogen column density derived from our spectral fits and the foreground Galactic value derived from radio maps (Stark et al. 1992) is $1.13 \cdot 10^{20}$ cm$^{-2}$ and the uncertainty in this mean is less than $0.10 \cdot 10^{20}$ cm$^{-2}$ (we compute the upper limit to the uncertainty in the mean conservatively using the 90 per cent confidence intervals as was done in Sect. 5.1). If we assume a Galactic dust-to-gas ratio this corresponds to a value of $E(B - V)$ (Bohlin, Savage & Drake 1978; Burstein & Heiles 1978) of less than 0.01, too small to explain the absence of broad Balmer lines of hydrogen by dust aborption provided the X-rays and optical flux travel to us via the same paths. Electron scattering of X-rays is constrained by the rapid soft X-ray variability described above but X-ray 'boreholes' through high column density gas such as a torus are possible. An extremely high dust-to-gas ratio of $\sim 20$ or more times the Galactic value would be needed to absorb the broad lines (and even then dust will provide significant X-ray photoelectic absorption for many grain size distributions). It is worth noting that Lawrence (1991) argues that much of the X-ray absorbing material in AGN is relatively dust free. The fact that some NLS1 have larger $E(B - V)$ values, deduced from the Balmer decrement, than expected based on their ROSAT columns might be due to the fact that a large amount of soft X-ray heating can move the Balmer decrement away from the standard case-B value toward a value more appropriate for pure collisional excitation (Gaskell & Ferland 1984).

## 6. Discussion: Models for NLS1 and relevant physical processes

### 6.1. Summary of NLS1 facts and discussion outline

Multiwavelength observations of NLS1 reveal that:

1. As a class NLS1 have generally steeper soft X-ray continua than normal Seyfert 1s. However, not all NLS1 are remarkably steep (cf. Sect. 4 and Table 1).
2. Many NLS1 have demonstrated rapid soft X-ray variability. PKS 0558−504, which satisfies the Goodrich (1989) criteria for classification as a NLS1, shows evidence for relativistic beaming based on its X-ray variability.
3. Two NLS1, IRAS 13224 − 3809 and RE J 1237+264, have shown *giant* amplitude soft X-ray variability.

4. The Balmer lines of hydrogen from NLS1 are only slightly broader than the forbidden lines such as [O III], [N II] and [S II].

5. NLS1 often have emission lines from Fe II or higher ionization iron lines such as [Fe VII] 6087 Å and [Fe X] 6375 Å. NLS1 generally have quite large Fe II to H$\beta$ ratios.

6. NLS1 do not appear to have dramatically larger Fe II EW values than normal Seyfert 1s, although some trend towards larger Fe II EW is possible (cf. Sect. IV of Goodrich 1989 and also compare the NLS1 with the normal Seyfert 1s in Fig. 1b of Joly 1991). This is somewhat in contrast to the much stronger relations found by Zheng & O'Brien (1990) in QSOs and Puchnarewicz et al. (1992) (see their Sect. 7.6.2) in ultrasoft AGN. Fe II lines are notoriously difficult to measure so this discrepancy may be due only to the different measurement techniques employed.

7. NLS1 have generally smaller H$\beta$ equivalent widths than normal Seyfert 1s (cf. Fig. 8 of Goodrich 1989).

8. Goodrich (1989) points out that in some NLS1 the H$\beta$ line appears to be more redshifted than [O III]. NLS1 where the H$\beta$ line appears to be less redshifted than [O III] have not been observed. Boroson & Green (1992) observed a similar trend in radio-loud QSOs but not in radio quiet ones (cf. their Sect. 4.3). Additionally, Fig. 5 of Boroson & Green (1992) shows that the QSOs that are the most similar to NLS1 in terms of their Fe II/H$\beta$ ratios tend to have asymmetric excess emission in the blue wings of their H$\beta$ lines.

9. Goodrich (1989) has shown that at least six NLS1 have significant mean continuum polarizations. He argues that the polarigenic mechanism is dust reflection based on the fact that the polarization rises to the blue as well as line broadening arguments (which are especially strong for Mrk 507). He presents some evidence which suggests that the scatterers are outside the BLR but inside or mixed with the NLR. Some NLS1 do not appear to have significant mean continuum polarizations.

10. Several NLS1 (e.g. I Zw 1, Kaz 163, RE J 1237+264, PKS 0558 − 504) have remarkably weak forbidden line spectra (Sect. III of Phillips 1977; Sect. V.b of Remillard et al. 1986; Sect. 2.8 of Jackson & Browne 1991; Brandt, Pounds & Fink 1995).

11. When compared to normal Seyfert 1s, some NLS1 are known to show deficits of ultraviolet emission relative to their soft X-ray spectra (Ark 564, Mrk 766, Fig. 8 of Walter & Fink 1993, Brandt et al. 1994) or entire spectral energy distributions (IRAS 13224 − 3809, NGC 4051, RE J 1034+393, Mas-Hesse et al. 1994, Sect. 2.5 of Puchnarewicz et al. 1995).

12. The 6-cm radio powers of NLS1 are within the range typical for other Seyfert galaxies (cf. Fig. 2 of Ulvestad, Antonucci & Goodrich 1995).

13. The median radio size of the VLA studied NLS1 is no larger than 300 pc, similar to the median radio size found for other Seyferts (Ulvestad & Wilson 1989; Ulvestad, Antonucci & Goodrich 1995).

14. Mrk 766 and Mrk 1126 have their radio axes oriented approximately perpendicular to their optical polarization position angles but 5C3.100 does not. Mrk 766 and Mrk 1126 are the *only* two known Seyfert 1s with radio axes oriented approximately perpendicular to their optical polarization position angles (Goodrich 1989; Ulvestad, Antonucci &

Goodrich 1995).

These are the properties that any model for NLS1 should be able to explain. It is worth noting that similar combinations of some of the properties of NLS1 have been seen in studies of different samples of objects. For example, Boroson & Green (1992) have found that QSOs with large Fe II/H$\beta$ and weak [O III] tend to have small H$\beta$ FWHM values. Others, such as perpendicular radio axes/optical polarization position angles combined with rapid soft X-ray variability and no evidence for large intrinsic cold absorption, so far appear to be peculiarly unique to NLS1 (e.g. Mrk 766). It must be remembered, as with Seyfert 1s in general, that NLS1 are probably a very heterogeneous class of objects and some properties above may be pathologies of individual objects rather than ailments of the class as a whole.

In the rest of this section we present models for NLS1 as well as physical processes that may be relevant to understanding them. We will always work under the assumption that NLS1 and normal Seyfert 1s are essentially the same class of object, as has been argued by Goodrich (1989). Diagrams such as our Fig. 8 and Fig. 8 of Goodrich (1989) suggest that NLS1 have extremal values of some underlying Seyfert 1 'control' parameter since they lie at the ends of distributions (cf. the 'phenomenological space' in Fig. 2 of Lawrence 1987). We will use the data to examine NLS1 models where NLS1 are Seyfert 1s with extremal values of orientation, black hole mass and/or accretion rate, warm absorption and BLR thickness. We will make the assumption that there is only one parameter which when extreme makes a Seyfert 1 a NLS1. This assumption must always be kept in mind when evaluating NLS1 models, since it has not been proven true (and indeed some of the properties described above hint at significant heterogeneity).

We will begin the discussion by describing possible relations between a steep ROSAT spectrum and the BLR. We will then model NLS1 as Seyfert 1s with extremal values of (1) orientation, (2) black hole mass and/or accretion rate, (3) warm absorption and (4) BLR thickness and confront these models with the facts listed above. We point out the main failures the models can plausibly explain and also point out their main failures. We then briefly discuss the implications if NLS1 do not show spectral breaks and their Fe II emission.

### 6.2. Steep ROSAT spectra and the BLR

The remarkably steep spectra we measure for NLS1 in the ROSAT band suggests that their EUV and X-ray spectral energy distributions may be somewhat different from those of normal Seyfert 1s. Since the formation and confinement of BLR clouds may depend on the EUV and X-ray spectra of AGN, it is worth considering the possibility that their unusual optical line properties and the steep ROSAT spectra we measure may be *directly related*. It is difficult to make quantitative predictions about this matter since almost nothing is known about the EUV and hard X-rays of NLS1 as well as BLR cloud formation and confinement. This lack of knowledge is particularly relevant for NLS1 since, as is discussed in Sect. II of Guilbert, Fabian & McCray (1983), steep X-ray spectra make the curve of gas temperature as a function of ionization parameter $\Xi$ very sensitive to the EUV spectrum because of the effects it has on ionization balance, line emission and photoelectric heating. Two extreme possibilities deserve elaboration:

1. The EUV and X-ray spectra of NLS1 are such that the double valued behaviour of the curve of gas temperature as a function of $\Xi$, often associated with multiphase BLR cloud pressure confinement, is removed (cf. Fig. 1 of Guilbert, Fabian & McCray 1983). Simple pressure confinement BLR models are probably incorrect but generally similar behaviour might occur in magnetically aided pressure confinement (Kallman & Mushotzky 1985), shock induced cloud formation (Perry & Dyson 1985) and bloated star BLR cloud (Scoville & Norman 1988, Alexander & Netzer 1994) models (cf. Brandt et al. 1994 for an example). It is worth noting that the $H\beta$ luminosity/$H\beta$ FWHM relation shown in Fig. 8 of Goodrich (1989) could be explained in the context of this model since steeper soft X-ray spectra lead both to fewer BLR clouds (and thus less $H\beta$ emission) as well as narrower lines.

2. NLS1 have hotter soft excesses that turn over at higher energies and thus the relative amount of flux in the UV and EUV bands is reduced (cf. Sect. 4.7). Fig. 1 of Guilbert, Fabian & McCray (1983) and the associated discussion point out that this will allow gas to remain cool at much higher X-ray illumination levels and can increase significantly the range of $\Xi$ over which gas temperature is double valued. Thus a BLR with greater physical and optical thickness may be formed (cf. Sect. 4.1 of Boroson & Green 1992).

### 6.3. NLS1 and orientation effects: the pole-on model

Relatively narrow NLS1 optical emission lines could arise via orientation effects. The gas motions in the BLRs of NLS1 might be confined to discs (cf. Wills & Browne 1986 and references therein) viewed in a more pole-on manner than in normal Seyfert 1s (e.g. Osterbrock & Pogge 1985; Stephens 1989; Puchnarewicz et al. 1992; Ulvestad, Antonucci & Goodrich 1995). Such motions would be expected if we viewed an extended accretion disc also in a pole-on manner, since otherwise clouds would collide with the disc.

If the pole-on model is correct and the BLR and accretion disc share the same axes of rotation, one would predict that NLS1 would show narrower X-ray emission lines from iron K$\alpha$ than normal Seyfert 1s due to the fact that the lines would be less broadened by Doppler and relativistic effects. This possibility might be tested with ASCA observations of the brightest NLS1 (e.g. Ark 564).

If NLS1 have unusually strong soft X-ray excesses rather than unusually weak hard X-ray tails (cf. Sect. 1) then the generally steep ROSAT slopes of NLS1 could arise via the angular dependences of the soft X-ray fluxes and albedos of accretion discs. Pole-on geometrically thick discs are expected to have brighter soft X-ray fluxes than edge-on geometrically thick discs (cf. Fig. 3 of Madau 1988) while the opposite holds for geometrically thin discs (Laor & Netzer 1989; cf. Netzer 1989 for a clear comparison of the angular dependences of geometrically thick versus geometrically thin discs). If the pole-on model for NLS1 could be independently proven true (and other potentially confusing effects such as black hole mass could be removed by, for example, dynamical and variability studies) this might offer a method for learning if the inner soft X-ray emitting parts of accretion discs are geometrically thick or geometrically thin (this point is discussed in Sect. 8.3.2.1 of Puchnarewicz et al. 1992). If some NLS1 have unusually weak hard X-ray tails rather than unusually strong soft X-ray excesses the situation becomes less clear. If the hard flux originates in a corona above a thin or thick disc, nonrelativistic geometrical effects would be fairly small for most geometries. If the hard flux originates in a corona located in the central cavity of a thick disc, it would be stronger, not weaker, in the pole-on scenario.

The pole-on model has trouble explaining the NLS1 with flat ROSAT spectra if steep NLS1 spectra are due to the angularly dependent emission and albedo from an accretion disc. It might, however, appeal to the intrinsic heterogeneity of Seyferts. NLS1 may have different sized and shaped discs due to, for example, different element abundances, accretion rates, specific angular momentum distributions and magnetic field strengths.

A pole-on orientation seems to be a natural one from which to observe relativistic beaming effects, since relativistic outflows are normally pictured to emerge from accretion disc axes, rather than from the sides of the discs. The extreme variabilities of PKS 0558 − 504 and IRAS 13224 − 3809 might be understood via beaming effects if these sources were viewed directly pole-on and we were looking down outflows from their central engines (although the interpretation of the soft flux becomes more difficult in this case). Giant amplitude variability might occur if directed emission passed in and out of our line of sight. The fact that NLS1 radio powers and extensions do not appear unusual, however, is at least empirically somewhat at odds with models with significant relativistic effects.

It is tempting to associate the optical line properties of NLS1 with pole-on orientation/anisotropy effects (e.g. Sect. 3.3 of Jackson & Browne 1991). One promising model in this respect postulates that the Fe II lines come from an accretion disc (cf. Collin-Souffrin 1988) while $H\beta$ and [O III] come from emission line clouds in a spherically symmetrical distribution. A pole-on orientation would show the greatest intensity in both the Fe II lines and the optical disc continuum, the first leading to a relatively larger Fe II EW, the second to relatively smaller $H\beta$ and [O III] EWs. Unfortunately, models of this type run into problems with Boroson (1992) and Boroson & Green (1992), at least for radio-quiet QSOs. Boroson (1992) has shown, given isotropy of the [O III] emission and that he has the same orientation distributions of radio-loud and radio-quiet objects in his UV excess selected sample, that the EW of [O III] is primarily a function of [O III] luminosity rather than an orientation dependence of the optical continuum. Furthermore, Boroson & Green (1992) point out that the fact that they obtained the same widths for the Fe II and $H\beta$ lines in their sample suggests that these lines arise in the same region and do not have strongly different anisotropies. These facts suggest that, if UV excess selected radio-quiet QSOs and NLS1 have similar orientation dependences (and indeed it is worth noting that some of the objects from Boroson & Green 1992 are NLS1), the pole-on model will require added complexity to explain the optical line strengths of NLS1.

Mrk 766 may have ionization cones and the pole-on model and some Seyfert 1 ionization cones models conflict (cf. Wilson 1995).

If the pole-on BLR interpretation is correct for NLS1, the fact that NLS1 host galaxies have a range of orientations would imply that the rotation axes of BLRs are not generally aligned with the rotation axes of their host galaxies. There are both observational and theoretical grounds for believing that this state of affairs is plausible (e.g. Tohline & Osterbrock 1982; Sect. V.a of Ulvestad & Wilson 1984; Miyoshi et al. 1995).

### 6.4. NLS1 and black hole masses, accretion rates and ultrasoft states

If the gravitational force from the central black hole is dominant in causing the motions of Seyfert BLR clouds (e.g. orbital motion, infall, or outflow at close to the escape velocity; Fig. 2 of Wandel & Yahil 1985 hints that gravitation, or at least some force that depends on the black hole mass, makes the BLR clouds move), narrower NLS1 optical emission lines will result from smaller black hole masses (due to smaller cloud Keplerian velocities) as long as the characteristic BLR distance from the central source does not change strongly with black hole mass. The BLR distance might not change strongly with black hole mass since the BLR consists of a photoionization dominated plasma and the luminosities of NLS1 and normal Seyfert 1s are generally comparable. Fuel limited accretion may set a typical Seyfert BLR density. As pointed out by Laor et al. (1994), a specific model where the characteristic BLR distance would not be a strong function of black hole mass is the dusty NLR model of Netzer & Laor (1993). In this model the radius of the BLR is set by the radius at which line supressing dust sublimes. The dust sublimation radius depends mainly on the source luminosity which, as noted above, is not strongly different for NLS1 and normal Seyfert 1s.

Smaller black hole masses could help to explain the rapid X-ray variability of NLS1 since the light crossing times and dynamical timescales of their central engines would be smaller.

NLS1 with smaller black holes would have to be accreting at higher fractions of their Eddington rates so as to maintain their relatively normal observed luminosities. This is plausible on energetic grounds and might be naturally explained if the speed at which the host galaxy can fuel the central monster is the limiting factor in accretion. Fuel limited accretion is plausible since matter in the surrounding galaxy and stellar cluster has $\sim 1000$ times or more the specific angular momentum of matter orbiting a supermassive black hole and cannot accrete until it loses this specific angular momentum (for convincing arguments that we do not understand the fueling process see Phinney 1994).

Viscous processes and X-ray reprocessing in optically thick matter near the central engine produce soft X-ray flux (Matsuoka et al. 1990, Pounds et al. 1990). Recent computations of accretion disc spectra (Ross, Fabian & Mineshige 1992; Shimura & Takahara 1993; Życki et al. 1994) show that the soft component should fall off around 0.5–1 keV and the illuminating power-law (with $\Gamma \approx 2$) will then be dominant. If NLS1 have stronger soft components or soft components that extend to higher energies than usual, this would make them appear to have steeper ROSAT spectra since there would be more flux in the low energy part of the ROSAT band. Smaller mass black holes are predicted to have hotter soft X-ray components that extend to higher energies (compare, for example, Fig. 3 and Fig. 4 of Ross, Fabian & Mineshige 1992). The resultant EUV and X-ray spectrum might then hinder BLR cloud formation near the central engine where cloud velocities would be highest (cf. Sect. 6.2). The work of Laor et al. (1994) suggests that at least some NLS1 could have weaker hard tails than usual rather than hotter soft excess components. If this is so, then a direct relationship between black hole mass (and/or accretion rate as a fraction of the Eddington rate) and hard tail strength would need to be found, and none are totally clear at present. One possibility, however, could invoke the potentially different dependences on black hole mass of electromagnetic energy extraction luminosity and accretion luminosity (cf. Sect. 5.2 of Bechtold et al. 1994; note this difference depends on how the magnetic field strength scales with black hole mass). If electromagnetically extracted energy were injected into a disc corona by magnetic fields, then smaller black hole masses might produce less corona (hard X-ray tail) luminosity relative to accretion (soft X-ray excess) luminosity.

The high Eddington fractions of NLS1 could make them the supermassive black hole analogues of the high ultrasoft states of Galactic black hole candidates. Galactic black hole candidates in their low states often have spectra that are qualitatively similar to those of normal Seyfert 1s. Occasionally, however, they make transitions to high states believed to be associated with increased accretion rates where a blackbody-like component becomes dominant below about 8 keV (e.g. Inoue 1992, Tanaka 1992). It would be an attractive unification of solar mass and supermassive black hole phenomena if NLS1 were AGN in ultrasoft high states. Ideas of this type have a long lineage but have never been applied to NLS1 as a class before (e.g. White, Fabian & Mushotzky 1984; Fiore & Elvis 1995). Of course, the accretion rate is the variable parameter in Galactic black hole candidates while in this NLS1 scenario it is the smaller mass rather than the larger accretion rate that leads to the ultrasoft behaviour. One apparent difficulty with this analogy is that Galactic black hole candidates tend to be less variable while in their ultrasoft states (their 'flickering' is reduced or removed, for example), while NLS1 do not show reduced variability.

The low black hole mass possibility discussed above is clearly only speculative currently and needs further examination so that it might be supported or falsified. ASCA observations of NLS1 will allow a high resolution study of their harder emission and will be able to establish whether the expected transition region between the soft excess and hard tail emission is seen. Implications if the spectrum remains steep above $\sim 2$ keV and there is no hard tail are discussed below. Simultaneous fitting of realistic accretion disc models to ROSAT, ASCA, and other X-ray and ultraviolet satellite data as well as X-ray variability power spectra analyses (e.g. Bao & Ostgaard 1994) will allow rough estimates of black hole masses. In addition, further systematic studies and temporal monitoring of the BLR and NLR spectral lines of NLS1 would be of use so that they might, for example, be plotted in a dynamical mass/X-ray variability diagram like that shown in Fig. 1 of Wandel & Mushotzky (1986). It is interesting that both Mrk 766 and NGC 4051 (cf. Sect. 5.1 and Bao & Ostgaard 1994) fall towards the low mass corner of their diagram. Dynamical masses for the NLS1 I Zw 1 ($9.3 \cdot 10^7$ solar masses), Mrk 1044 ($1.1 \cdot 10^7$ solar masses) and Mrk 896 ($1.3 \cdot 10^7$ solar masses) are given in Padovani & Rafanelli (1988). These fall towards the low end of Fig. 5 of Padovani (1989), but there are still some non-NLS1 objects that appear to have smaller dynamical masses. The NLR lines of NLS1 may also be narrow relative to the NLR lines of normal Seyfert 1s if the tentative low mass black hole idea is correct, but this is by no means a certainty. First of all, gravitation would have to play the dominant role in accelerating the clouds of the NLR (this is generally supported by Whittle 1992a). Secondly, the gravita-tion must be provided predominantly by the black hole rather than by other masses. AGN may have central star clusters with masses that can substantially exceed the mass of the central hole for extended periods of time (e.g. Murphy, Cohn and Durisen 1991). The mass of the torus may also be comparable to the mass of the central hole in some cases (cf. Planesas et al. 1991). Since $r_{BLR} \ll r_{Cluster} \ll r_{NLR}$, the dynamics of the NLR may be dominated then by the gravitational pull from the central star cluster rather from the black hole. The central stellar cluster masses in galaxies with different formation/evolutionary/merger/tidal interaction histories need not be simply related to the masses of their central black holes (for a glimmer of hope that the stellar cluster and black hole mass are proportional see Sect. 4b of Wandel & Mushotzky 1986 but then see Whittle 1992b for known complicating effects). Thirdly, the NLR will also have to have a size which does not depend strongly on the mass of the central black hole.

It is also possible that the ultrasoft spectra of NLS1 are due to increased accretion rates and that NLS1 have typical Seyfert 1 black hole masses. Pounds (1994) has discussed this possibility for the ROSAT Wide Field Camera AGN RE J 1034+393 but not for NLS1 as a class. This scenario would make NLS1 an even more direct analogy of the ultrasoft Galactic black hole candidates. In this case, with extended X-ray monitoring of many Seyfert 1s, one might expect to see transitions from the ultrasoft NLS1 state to the normal Seyfert 1 state or vice versa since such transitions have been seen in Galactic black hole candidates. If the transitions were long term as they are in Galactic black hole candidates, optical monitoring of the broad lines would allow tests of BLR models. The simple pressure confined BLR model, for example, predicts that the subsequent dissipation (if a transition to a NLS1 state were made) or formation (if a transition to a normal Seyfert 1 state were made) of BLR clouds would take $\sim 100$ yr (cf. equation 9 of Guilbert, Fabian & McCray 1983).

The fact that some NLS1 do not appear to have especially steep spectra presents a problem for the low mass black hole model as well. This might be explained within the hindered BLR mechanism if NLS1 have spectra that vary in time about some mean steep spectrum since the BLR properties will depend on the averaged $\sim 100$ year behaviour of the illuminating spectrum and not necessarily on the spectrum observed at present (cf. equation 9 of Guilbert, Fabian & McCray 1983). Spectral variabilty could arise from fluctuations in the feeding rate of the central monster or via magnetic instabilities in a disc dynamo.

The $\sim 10^{47}$ erg s$^{-1}$ luminosity of the hot ($\sim 120$ eV) big blue bump in E1346+266 presents serious problems for a low mass black hole interpretation for this object since Eddington limit constraints suggest a black hole mass of $> 10^9$ solar masses.

### 6.5. NLS1 and warm absorption

Steep NLS1 ROSAT spectra could be formed if much of their hard fluxes were absorbed by warm gas along the line of sight (the presence of cold gas along the line of sight has already been constrained in Sect. 5.3). In warm gas where the lighter elements but not metals such as oxygen and iron have been fully stripped of their electrons, the opacity below $\sim 0.7$ keV can be greatly decreased while still allowing for absorption at higher energies. A model of this type might lead to weaker than usual hard X-ray tails (at least in the ROSAT band) relative to the near IR and optical continua (cf. Sect. 5.1).

A prediction of NLS1 models with large amounts of warm absorption is the imprinting of absorption edges on NLS1 X-ray spectra. The fact that models with edges fit some NLS1 such as Ark 564 (Brandt et al. 1994) and NGC 4051 (Mihara et al. 1994) well suggests that warm absorbers may be relevant in at least some cases (although it should be noted that even when warm absorption is taken into account both Ark 564 and NGC 4051 still show evidence for steeper than usual soft spectra). However, it seems unlikely that the steep ROSAT spectra of all NLS1 can be explained via warm absorption. As described in Sect. 5.1, for example, analyses of the spectral energy distributions of E1346+266, RE J 1034+393 and Mrk 478 suggest that they have truly large soft excesses rather than weakened hard tails. Additionally, RE J 1237+264 has shown variability by a factor of $\sim 70$ in ROSAT count rate without any evidence for a change in spectral shape. If this variability arose via changes in the ionization parameter and/or column density of a warm absorber a strong change in spectral shape would generally be expected. If it instead arose via an intrinsic change in central source soft X-ray luminosity a change in spectral shape would still be expected for a warm absorber where photoionization plays a role in determining its ionization structure (even if it were not in strict photoionization equilibrium). Molendi & Maccacaro (1994) have argued that variations in the optical depth and/or ionization state of a warm absorber have difficulty explaining the spectral variability of Mrk 766. Mrk 335, which lies among the NLS1 in Fig. 7, shows a large soft excess yet no evidence for low energy sharp spectral features in the preliminary analysis of a recent ASCA observation (P. Życki, pers. comm.).

Warm absorbers with internal dust can explain many NLS1 properties but do not appear to naturally explain their flux or spectral variability (cf. Brandt, Fabian & Pounds 1995).

### 6.6. NLS1 and BLR thickness/density

Boroson & Green (1992) have proposed a model for the Fe II and [O III] anticorrelation seen in QSOs whereby a toroidal distribution of BLR clouds screens a roughly planar NLR from central source photons (cf. their Sect. 4.1). Sources with stronger Fe II and weaker [O III] are posited to have thicker or denser (cf. Gaskell 1985) toroidal distributions of clouds which serve as stronger Fe II sources as well as more effective screeners of central source photons. We shall consider the application of this model to NLS1.

If the formation and confinement of BLR clouds depends on the EUV and X-ray spectra of NLS1, then the steeper than usual ROSAT slopes of NLS1 could be indicative of EUV and X-ray spectra that lead to thicker BLRs (cf. Sect 6.2; although note that models of this type may have difficulty with Sect. III of Gaskell 1985). If, on the other hand, the survival of the BLR clouds does not depend on the form of the EUV and X-ray spectra then the very clear tendency we see for steeper soft X-ray spectra in NLS1 must be explained another way within this model if it is going to be considered a complete description of them. In addition the smaller H$\beta$ FWHM values and other NLS1 properties of Sect. 6.1 require explanation.

A low black hole mass in conjunction with a thicker or denser BLR (perhaps but not necesarily induced by the EUV and X-ray spectrum of a smaller mass hole) explains many of the properties of NLS1. In particular such a model would enjoy both the advantages of smaller mass black holes described in Sect. 6.4 as well as the advantages of the Boroson & Green (1992) model described above. One constraint we have on this model is that the soft X-ray flux we see does not show evidence for large amounts of cold absorption. Thus the soft X-ray photons we see must travel to us primarily through the space between the BLR clouds rather than through the clouds themselves. This would naturally be the case if the BLR clouds were quite optically thick. The smaller H$\beta$ equivalent widths of NLS1 could arise in thick clouds since the 'beaming factor' of H$\beta$ back toward the ionizing source will be large due to its relatively large line opacity (cf. Fig. 6 of Ferland et al. 1992). Fe II will be much less beamed back toward the central source due to its generally smaller line opacity.

One problem with the Boroson & Green (1992) model is that observations of AGN generally indicate that their NLRs do not have planar geometries that might have ionizing photons screened from them by a thicker or denser than usual toroidal distribution of BLR clouds. In particular some NLR [O III] emission is often observed to lie along the radio axes in a roughly conical shape while some is more widely distributed (e.g. Whittle et al. 1988; Macchetto et al. 1994). Modifications of the Boroson & Green (1992) geometry where the constraints of a toroidal distribution of BLR clouds and a planar NLR are relaxed, but where a thicker or denser BLR still prevents photons from reaching the NLR, are perhaps possible. Of course, both the torus and the BLR clouds may be enhanced, and this would then echo the suggestion by Ulvestad, Antonucci & Goodrich (1995) that NLS1 (and not all Seyfert 1s) are the pole-on equivalents of Seyfert 2s. It is relevant that Heckman et al. (1989) have found from the CO and far-infrared properties of Seyferts that Seyfert 2s have intrinsically more dust and molecular gas than Seyfert 1s. It would be worthwhile to compare these properties in NLS1 and Seyfert 2s to see if they are more consistent with each other. The warm absorbers and polarization by dust scattering seen in some NLS1 might naturally be associated with a generally thicker or denser gaseous environment.

If NLS1 have thicker or denser BLRs but their BLR clouds are not primarily accelerated by gravity, radiative acceleration models may be relevant. Gaskell (1985) has demonstrated that the H$\beta$ EW versus H$\beta$ FWHM correlation can be explained in the context of radiative accelereration models. Steep soft X-ray spectra will affect radiative acceleration, perhaps leading to the excess H$\beta$ redshifts seen.

### 6.7. Implications if NLS1 show no spectral breaks

The NLS1 that have steep spectra in the ROSAT band have not been well studied above 2.4 keV. Thus the possibility that some or all of their spectra remain steep up to 10 keV and beyond cannot currently be excluded. Indeed, using ROSAT and EXOSAT data Gondhalekar et al. (1994) find no evidence for a break below $\sim 4$ keV in Mrk 478. In the absence of spectral breaks, a 2–10 keV photon index of $\approx 3$ would have important consequences for the nonlinear nonthermal electron positron pair cascade models that are often invoked to explain the ultraviolet to gamma-ray spectra of gamma-ray weak AGN (e.g. Zdziarski 1992, Svensson 1994). Saturated pair cascade models naturally predict an underlying photon index of about 1.9 and the absence of a break from $\Gamma \approx 3$–4 to $\Gamma \approx 1.9$ in the 2–10 keV range would present difficulties for such models that would require potentially unattractive changes in the hard compactness parameter, the injection function or both.

### 6.8. NLS1 and Fe II line formation

If the observed ROSAT continuum is extrapolated to hard X-rays, with the observed slope, one finds an extremely weak hard X-ray continuum. The reprocessing model predicts that in such a case the optical Fe II lines must be weak, contrary to the observations. The trend of having stronger Fe II lines in steeper X-ray continuum sources (e.g. Shastri et al. 1993) holds for the somewhat extreme case of NLS1. If the reprocessing model is correct, we expect a flattening of the X-ray continuum beyond the ROSAT PSPC band to provide enough hard X-ray energy to power the Fe II lines. Harder X-ray observations thus provide a key to check the reprocessing model for Fe II lines.

### 7. Summary

Narrow-line Seyfert 1 galaxies (NLS1), with hydrogen Balmer lines with FWHM in the range $\approx 500$–1500 km s$^{-1}$, have generally steeper 0.1–2.4 keV continuum slopes than normal Seyfert 1s (see Fig. 8). Simple power-law model fits to the data yield photon indices *often above 4*. We detect rapid soft X-ray variability in many NLS1 and use this variability to rule out the possibility that most of the soft X-rays in NLS1 are scattered into our line of sight by electrons in a scattering mirror. Our fits do not show large amounts of soft X-ray cold absorption, and thus simple obscuration by cold matter along the line of sight cannot easily explain the narrow hydrogen Balmer lines of NLS1.

Steep ROSAT spectra suggest that the EUV and X-ray spectral energy distributions of NLS1 may be somewhat different than those of normal Seyfert 1s, and these different spectral energy distributions can strongly influence BLR cloud formation and confinement. The large soft X-ray excesses seen in at least some NLS1 probably cannot be produced by the reprocessing of hard flux due to the fact that for reasonable models of the hard flux there is more luminosity in the soft component than in the hard one. The X-ray steep NLS1 have not been well studied above 2.4 keV and some of their spectra could remain steep at higher energies. This would have important consequences for saturated nonthermal electron positron pair cascade models as well as reprocessing models for the large Fe II/H$\beta$ ratios of NLS1.

We test models for NLS1 where they are Seyfert 1s with extremal values of pole-on orientation, black hole mass and/or accretion rate, warm absorption and BLR thickness using our new data. All simple models appear to have drawbacks. Models with smaller mass black holes as well as thicker BLRs show some promise, perhaps in accordance with suggestions that only NLS1 and not all Seyfert 1s are the pole-on equivalents of Seyfert 2s. We suggest specific further tests of NLS1 models.

NLS1 may be analogous to the high ultrasoft states of Galactic black hole candidates.

*Acknowledgements.*

We thank Prof. Joachim Trümper for important suggestions for improvements. We thank Andy Fabian, Bob Goodrich, Tsuneo Kii, Ari Laor, Giorgio Matt, Richard Mushotzky, Paul Nandra, Hagai Netzer, Chiko Otani, Ken Pounds, Liz Puchnarewicz and Chris Reynolds for useful discussions. We thank the anonymous referee for a thorough proofreading and helpful suggestions. WNB gratefully acknowledges financial support from the United States National Science Foundation and the British Overseas Research Studentship Programme. The ROSAT project is supported by the Bundesministerium für Forschung und Technologie (BMFT) and the Max-Planck Society. This research has made use of the NASA/IPAC extragalactic data base which is operated by the Jet Propulsion Laboratory, Caltech.